\documentclass{psp-book975x65}
\usepackage{psp-book-van}          
\usepackage{subfigure}
\usepackage[catalan,english]{babel}
\usepackage[T1]{fontenc}		
\copyline{Book Title}{2009}{978-981-nnnn-nn-n}
\title{My Book Title}              

\input{Commands.tes}
\graphicspath{{/Figures/}}

\begin{document}
\DeclareGraphicsRule{.eps.gz}{eps}{.eps.bb}{` gunzip -c #1}


\setcounter{page}{1}

\chapter{Time dependent phenomena in nanoparticle assemblies} 
\label{Interac_Chap}
\author{Òscar Iglesias}
\author{\it Departament de F\'{\i}sica Fonamental and IN$^2$UB,
Universitat de Barcelona, Diagonal 647, 08028 Barcelona, Spain}
\texttt{oscar@ffn.ub.es}

Magnetic nanoparticle systems are difficult systems to model due to the interplay between intrinsic and collective effects. The first ones are associated to the magnetic properties of an individual particle and require considering the atomic spins of the magnetic ions, their mutual exchange interactions and magnetocrytalline anisotropy. Due to its finite size, an individual particle has different magnetic properties from the bulk counterpart material. Morover, the high proportion of surface spins with reduced coordination influences the equilibrium magnetic configuration of individual particle, with spin noncollinearities being a consequence of the distinct surface anisotropy. The collective effects have to do with interactions among the nanoparticles in an ensemble such as long range dipole-dipole interactions and can be tackled more easily by considering each nanoparticle as one effective spin having the magnitude of the total magnization of the particle. Therefore, at this level of description, nanoparticle ensembles can be modeled by a collection of macrospins having their own anisotropy axes and interacting through dipolar interactions, neglecting their internal structure, which inturn is equivalent to assuming that the interactomic exchange coupling is strong enough to keep atomic magnetic ions aligned along the global magnetization direction. 
Whereas in atomic magnetic materials the exchange interaction usually dominates over dipolar interactions, the opposite happens in many nanoscale particle or clustered magnetic systems, for which the interparticle interactions are mainly of dipolar origin. Therefore, one spin models (OSP) should, in principle, provide a correct description of non-interacting systems and, to a first approximation be valid also to account for the main features observed in more concentrated samples where interactions cannot be neglected. It has been shown also recently \cite{KachachiGaranin_prl03,Yanes_prb07} that spin non-collinearities due to surface anisotropy can even be incorporated within the OSP approach if an effective cubic anisotropy term is added to the original uniaxial anisotropy energy. However, incorporation of dipolar interactions along these lines does not seem feasable within the present theoretical frameworks. 

While we have a valid theoretical framework to compute equilibrium magnetic properties (such as thermal dependence M(T), isothermal field dependence M(H), low temperature configurations,...) analytically or numerically within  the scope of OSP models \cite{GarciaPaladv00} for non-interacting systems, models including dipolar interactions can only compute these quantities using perturbative thermodynamic theory and, even so, anaytical expressions can only be obtained under certain limits and approximations. In contrast, dynamic properties (such as hysteresis loops, FC-ZFC processes, susceptibility or magnetic relaxation) are non-equilibrium phenomena for which a unique theoretical framework covering the wide range of time scales involved is not available even for non-interacting systems. Therefore, most studies on dynamics revert to numerical simulations of ensembles of macrospins (\cite{GarciaOteroprl00,Ulrichprb03,Russprb06,Serantesprb09,Fernandezprb00,Fernandezprb09,Alonsoprb10,Kechrakosprb98}) based on Monte Carlo (MC) methods since simulations based on the Landau-Lifschitz equation cannot access the long time scales involved. 

The main difficulty in modeling the long-time dynamics of magnetic NP ensembles is the calculation of the relaxation rates between metastable states as they depend on the energy barriers that has to be overcome by thermal fluctuations and, consequently, they depend on the orientation of the NP easy-axis with respect to the field axis. At the same time, the presence of interparticle interactions modifies in a complex manner the energy landscape due to the long range character of the dipolar interactions and several escape paths out of a metastable mininimum may be coexist. Therefore, in general, the energy barriers responsible for the thermal relaxation of the NP ensemble towards equilibrium cannnot be computed analytically and numerical simulations have to be used.

While dilute systems are well understood, experimental results for dense systems are still a matter of controversy. Some of their peculiar magnetic properties have been attributed to dipolar interactions although many of the issues are still controversial. Different experimental results measuring the same physical quantities give contradictory results and theoretical explanations are many times inconclusive or unclear.    
In the following, we briefly outline the main subjects to be clarified:
\begin{itemlist}
\item The complexity of dipolar interactions and the frustration provided by the randomness in particle positions and anisotropy axes directions present in highly concentrated ferrofluids seem enough ingredients to create a collective glassy dynamics in these kind of systems. Experiments probing the relaxation of the thermoremanent magnetization \cite{Jonssonprl95,Mamiyaprl98,Jonssonprl98}
have evidenced magnetic aging. Studies of the dynamic and nonlinear susceptibilities \cite{Jonssonprl98,Djurbergprl97,Kleemannprb01}, also find evidence of a critical behaviour typical of a spin-glass-like freezing. 
All these studies have attributed this collective SG behaviour to dipolar interactions, although surface exchange may also be at the origin of this phenomenon. However, MC simulations of a system of interacting monodomain particles \cite{GarciaOteroprl00} show that, while the dependence of ZFC/FC curves on interaction and cooling rate are reminiscent of a spin glass transition at $T_B$, the relaxational behaviour is not in accordance with the picture of cooperative freezing.

\item It is still not clear what is the dependence of the blocking temperature and remanent magnetization with concentration, $\varepsilon$, in ferrofluids: while most experiments \cite{Dormannjpc88,Gangopadhyayie93,Chantrellbook97,Chantrellie91,Elhilojm92,OGradyie93,Luoprl91,GarciaOteroprl00} find an increase of $T_B$ and a decrease of $M_R$ with $\varepsilon$, others \cite{Morupprl94,Morupprb95} observe the contrary variation in similar systems.

\item The dependence of the effective energy barriers with concentration is unclear. While theoretical studies by the group of Dormann \cite{Dormannjpc88,Dormannadv97} and experimental results by Luis \cite{Luisprb02} on Co clusters predict an increase of the barrier for magnetization reversal , Mörup \cite{Morupprl94} et al. argue for a reduction of $T_B$ with $\varepsilon$.

\item Although magnetic hysteresis is usually reported in experimental studies of nanoparticle ensembles for a wide range of sizes and concentrations, and varied compositions, a general theoretical framework able to account for the observed phenomenology is still lacking. However, we can say that, in general, for disordered systems, the dipolar interaction diminishes the coercive field and decreases the remanent magnetization \cite{Gangopadhyayie93,Chantrellbook97}.
\end{itemlist}

In this contribution, we will present a review of our works on the time dependence of magnetization in nanoparticle systems starting from non-interacting systems, presenting a general theoretical framework for the analysis of relaxation curves which is based on the so-called $\svar$ scaling method. We will detail the basics and explain its range of validity, showing also its application in experimental measurements of magnetic relaxation. We will also discuss how it can be applied to determine the energy barrier distributions responsible for the relaxation. Next, we will show how the proposed methodology can be extended to include dipolar interactions between the nanoparticles. A thorough presentation of the method will be presented as exemplified for a 1D chain of interacting spins, with emphasis put on showing the microscopic origin of the observed macroscopic time dependence of the magnetization. Experimental application examples will be given showing that the validity of the method is not limited to 1D case.

\section{Magnetic relaxation in non-interacting NP ensembles}
\label{Nonint}
Let us consider a general magnetic system, not necessarily a system of small particles. We will only assume that, whatever is the underlying microscopic model used to describe it, it can be thought in terms of effective energy barriers that separate the metastable states of the appropriate degree of freedom of the constituents. 
Therefore, we will be thinking now of a magnetic system as a collection of energy barriers $E$ that can be characterized by a certain distribution function $f(E)$ which contains the specific composition of the system. We are interested in the time dependence of the order parameter $m(t)$, which we will call magnetization thinking in applications to small particle systems.
In particular, in single-domain particles and granular materials the energy barriers due to the anisotropy are, in principle, proportional to the volume of the particle or grain. In this case $f(E)$ reflects the scattering of particle volumes or anisotropy constants. It should be noted that using a distribution of energies or of relaxation times is better and more general than using a distribution of volumes or particle sizes since in this way no assumption about the relation between these parameters has to be made. 

The decay of the magnetization of a distribution of single-domain particles is given by the relaxation law:
\begin{equation}
\label{Mt}
m(t)=\int_{0}^{\infty}dE\ f(E)\ e^{-t/\tau(E)}\ .
\end{equation}
where $f(E)$ is the distribution function of energy barriers that have to be overcome by thermal fluctuations, in order to change the equilibrium magnetization direction of the particles. The exponential factor is the classical Boltzmann probability for  a particle to change its equilibrium magnetization value, and $\tau (E)$ is the relaxation time used in Ne\'el's theory \cite{Neel}, given by:
\begin{equation}
\tau (E)=\tau_0\ e^{E/k_B T}\ ,  
\end{equation}
where 1/$\tau_0$ is an attempt frequency of the order $10^8$ - $10^{12}$ s$^{-1}$, $k_B$ the Boltzmann constant and $T$ the temperature.

Let us introduce the function $p(t,E)$ defined by 
\begin{equation}
p(t,E)=e^{-(t/\tau _0)\exp (-E/k_{B}T)} \ . 
\label{Snake}
\end{equation}
Taking into account that $p(t,E)$, for a given time $t$ varies abruptly from 0 to 1, as the energy barrier $E$ increases, the usual simplification \cite{Street} consists on approximating p(t, E) by a step function whose discontinuity $E_C(t)$ moves to higher values of $E$ as time elapses. As a consequence, the integral is cutoff at the lower limit by the value of $E_C(t)$, which is the only time-dependent parameter, and the expression \ref{Mt} is approximated by \cite{Labartaprb93}
\begin{equation}
\label{Mcrit}
m(t)\simeq \int_{E_c(t)}^\infty dE\ f(E)\ .  
\end{equation}
$E_c(t)$ corresponds to the energy barrier value for which the function $p(t,E)$ has the inflection point and is given by 
\begin{equation}
\label{Ec}
E_c=k_{B}T\ln (t/\tau _0).  
\end{equation} 
From Eq. (\ref{Ec}), we conclude that the remanent magnetization $M(t)$ obtained after integration over the energy barriers $E$ is a function of the parameter $E_c(t)=k_{B}T\ln (t/\tau_0)$. The existence of this scaling variable implies that measuring the magnetization as a function of the temperature at a given time is equivalent to measure the magnetization as a function of the $\ln (t)$ at a fixed temperature. This time-temperature correspondence is characteristic of activated processes governed by the Arrhenius law.

Moreover, in \refcite{Iglesiaszpb96}, we demostrated that the logarithmic relaxation rate $S(t)=\frac{\partial M(t)}{\partial (\ln (t))}$(also called magnetic viscosity) is related to the energy barrier distribution through the expression:
\begin{equation}
\label{Viscodis}
S(t)= k_B T f(\bar E_C) (1+S^{(1)}+S^{(2)}+\dots) \ ,
\end{equation}
where $S^{(n)}$ is proportional to $(k_B T/\sigma )^n$, $\sigma$ being the characteristic width of the energy barrier distribution. Therefore, at low enough temperatures, the corrections introduced by the $S^{(n)}$ terms can be neglected and the magnetic viscosity becomes directly proportional to the distribution of energy barriers. 

Experiments on magnetic relaxation are limited to a range of at most four decades in time, but during this range of times the magnetization of most physical systems only varies in a small percentage of the initial value, so that the range of energy barriers explored during the experiment is limited to a small fraction of the real distribution $f(E)$. This is so because of the spread of the physical properties of the systems and the exponential variation of the relaxation times with the energy. So, it would be interesting to find a method to extend the experimental relaxation curves to much longer times without having to perform impossibly long measurements. This is what the phenomenological $\svar$ scaling pretends. 

The method relies on the fact that under certain conditions there is a natural scaling variable in the relaxation law that relates temporal to temperature scales, thus making possible to deduce relaxation curves at long times and a given temperature from the knowledge of the short time relaxations at different higher temperatures.
The idea ressembles that found in earlier works by Pr\'ejean et al. in the context of relaxation in spin glasses at the beginning of the 80's \cite{Prejeanjpe80}.
\begin{figure}[htbp]
\centering
\includegraphics[width=1.0\textwidth]{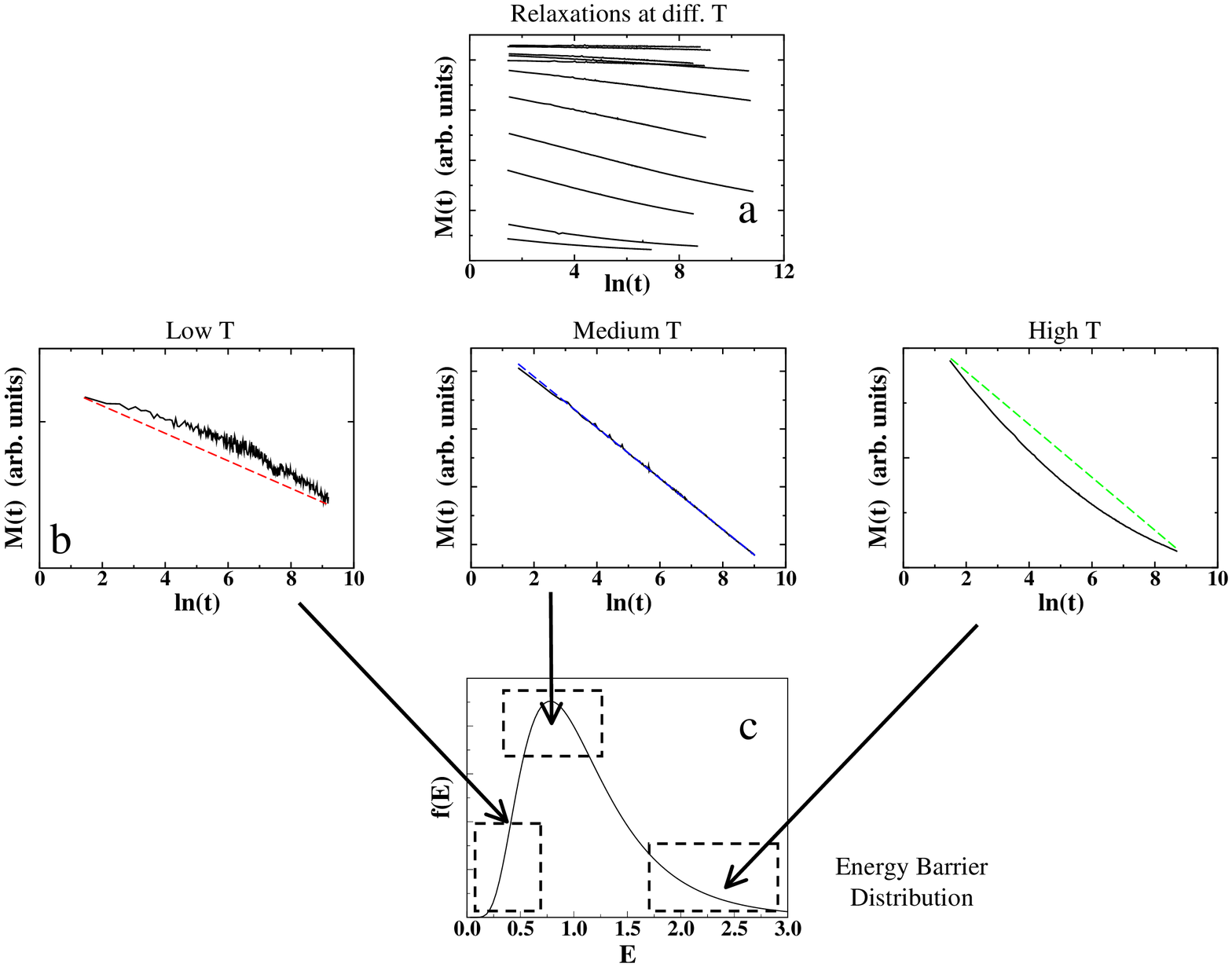}
\caption{Illustration of the origin of the $\svar$ scaling for relaxation curves at different $T$. (a) Relaxations curves at different $T$ (the lowest $T$ is for the uppermost curve) in logarithmic time scale as measured in an experiment. (b) Details of a low, medium and high $T$ curves. The dashed lines corrspond to logarithmic time dependence, departure from this law are clearly seen at low and high $T$. (c) Diagram showing the fraction of energy barriers that contribute to the above relaxation curves.}
\label{Scaling_fig}
\end{figure}

In order to illustrate its origin, let us start from a real experimental example. Let us consider a set of relaxation curves measured at different temperatures as the ones displayed in Fig. \ref{Scaling_fig}a. At a first glance, they all seem to follow a logarithmic law. One is even tempted to make linear fits which probably would be accurate enough, and that is what is usually done in the literature \cite{Sappeyjm00}. But let us take a closer inspection to every particular curve. In Figs. \ref{Scaling_fig}b we have selected a low, intermediate and high temperature curves and plotted them separately. It can be clearly seen that only at intermediate temperatures the curves are straight lines corresponding to a logarithmic law, but that, at high and low $T$, they are curved downward and upward respectively, showing a clear deviation from a logarithmic law. 

In order to understand this change of curvatures, let us notice that during the time of the experiment only a small fraction of all the energy barriers of the system are sampled. Moreover, this fraction is not the same at different temperatures. The typical energy barriers explored during the measuring time of the experiment, $t_m$, are of the order of $E_m=T\ln(t_m/\tau_0)$, proportional to the temperature, so that relaxation curves at different temperatures are not directly comparable because they collect results from different portions of the energy barrier distribution of the system. As Fig. \ref{Scaling_fig}c illustrates, the low temperature curve samples the low energy part of the energy barrier distribution $f(E)$, at intermediate temperatures the mean energy barriers are explored, and at high temperatures the high energy part of $f(E)$ is sampled. Moreover, this explains why only relaxations at intermediate temperatures are logarithmic, since, in this range, the fraction of barriers explored are near the maximum of the distribution where $f(E)$ is almost flat. The curvature of the relaxation curves is, therefore, directly related to the curvature of the energy distribution. This clearly indicates that, even though all the measurements have been performed during the same time window, the energy barriers responsible for the relaxation are not the same at every temperature.

To verify the validity of the $\svar$ scaling law in real small particles systems, a sample was cooled from above the blocking to the measurement temperature under a magnetic field and the magnetic relaxation data was measured after switching off the field and subsequently analyzed within the scope of the scaling hypothesis \cite{Labartaprb93}. 
According to the scaling hypothesis discussed previously,  all the different curves corresponding to different temperatures, would have to scale onto one single master curve when plotted as a function of the scaling variable $\svar$. In order to verify the validity of this model, we try to scale the relaxation data of the referred samples. The procedure used for this purpose consists on plotting the relaxation curves in a $M$ vs $Ln(t)$ plot and trying to connect each of them continuously with the adjoining curves corresponding to the nearest measured temperatures. To do that, we shifted the experimental curves in the $T\ln (t)$ axis by an amount equal to $T\ln (\tau_0)$, where $\tau_0$ is a characteristic time which governs the relaxation processes on an atomic scale. $\tau_0$ is the same for all of the measured temperatures and it was chosen to be the best in bringing all the curves into one. 
\begin{figure}[htbp]
\centering
\includegraphics[width=0.49\textwidth]{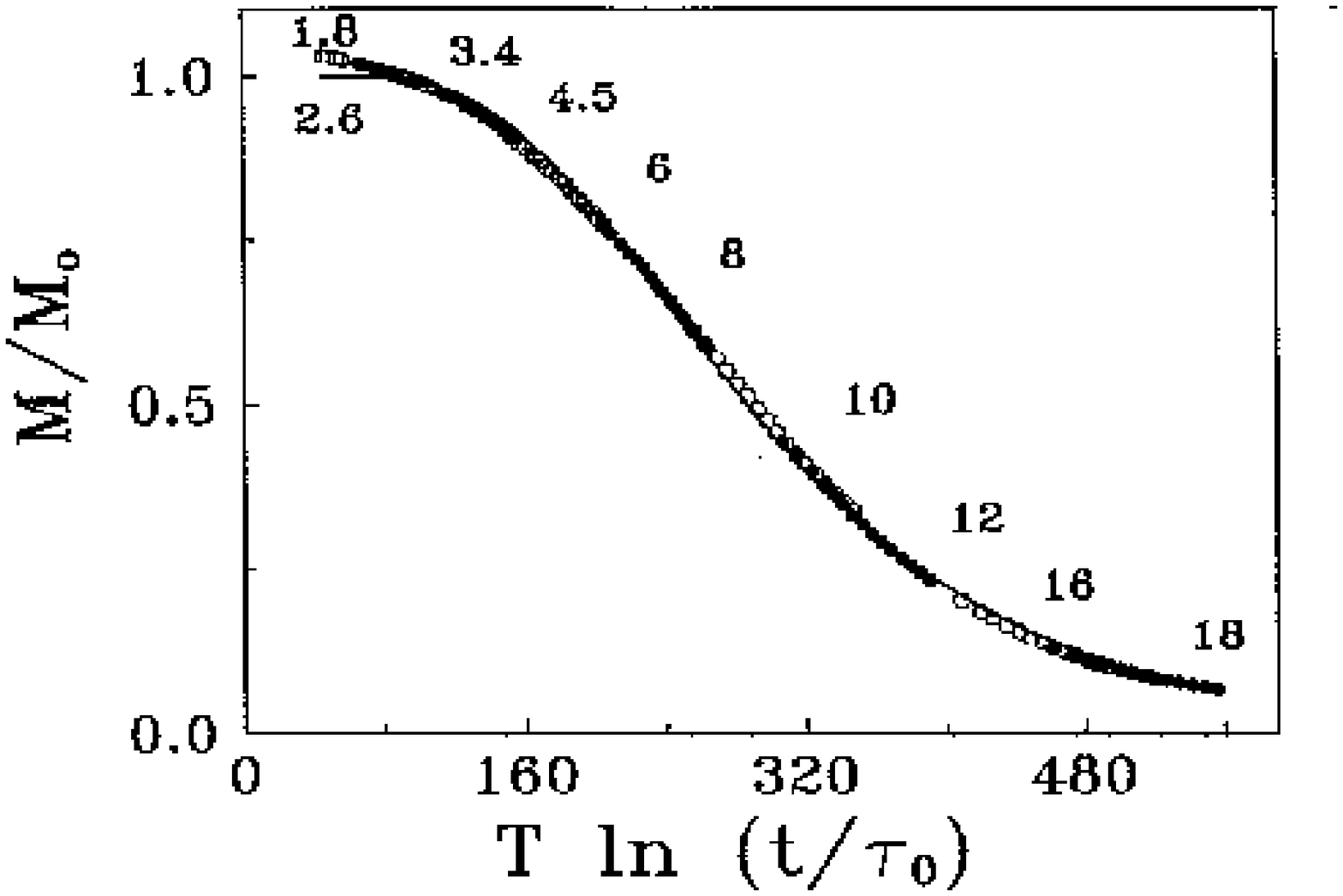}
\includegraphics[width=0.49\textwidth]{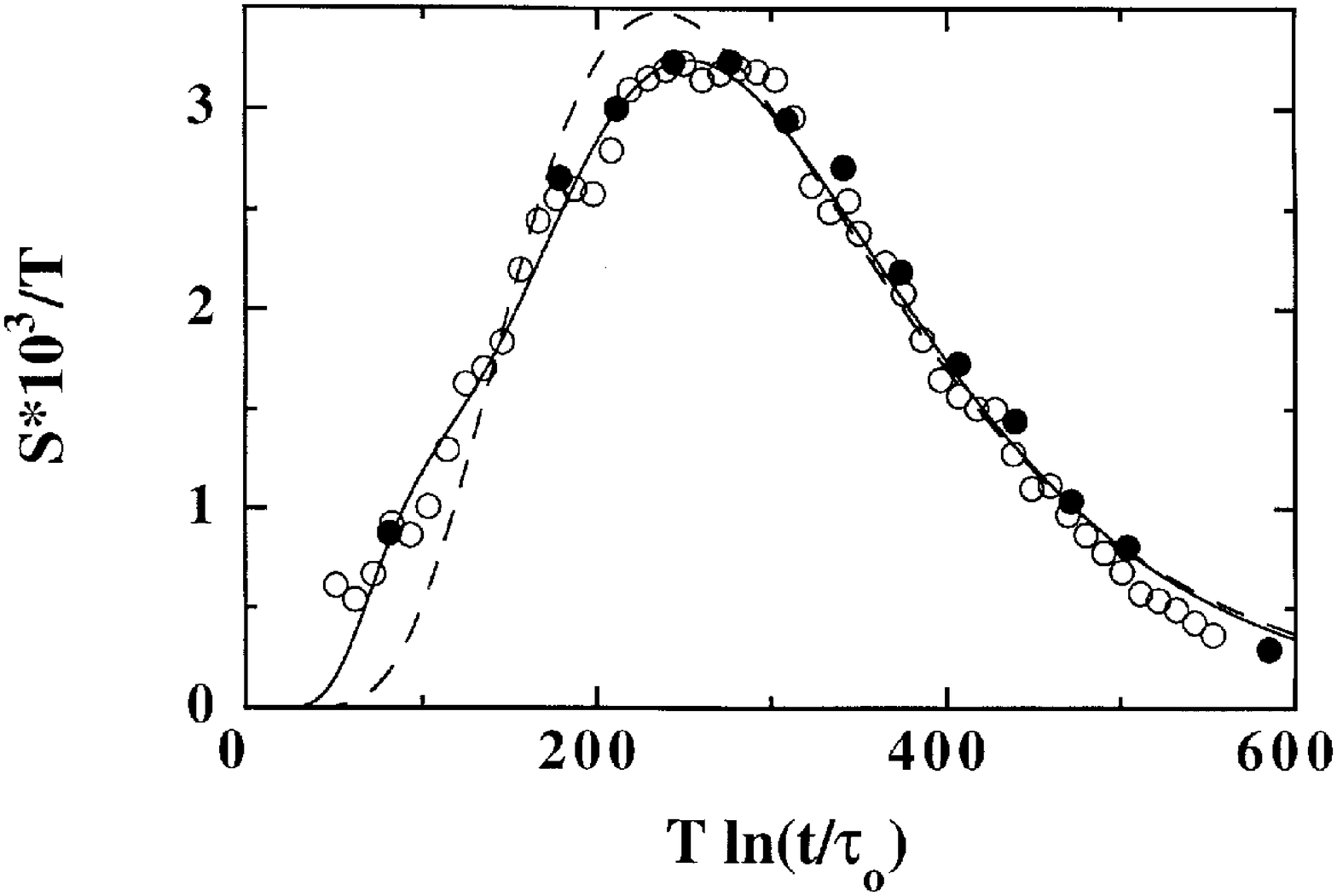}
\caption{Left panel: Master curve obtained by $\svar$ scaling of the relaxation curves for a ferrofluid composed of FeC particles. The figure shows the reduced magnetization as a function of the scaling variable. The corresponding temperatures are indicated above the corresponding interval. The solid line is the theoretical curve calculated by using Eq. (\ref{Mt}).
Right panel: Numerical derivative of the master curve with respect to the scaling  variable (open circles) and the energy distribution functions as obtained by fitting the master relaxation curve to expression (\ref{Mt}) assuming a 
single logarithmic-normal distribution (dashed curve) and two logarithmic-normal
distributions.The differential of the thermoremanence relative to the saturation magnetization versus the temperature is also shown in full circles for comparison.
}
\label{Tlntsam1_fig}
\end{figure}

In Fig. \ref{Tlntsam1_fig} the results of this scaling is shown. One of the most interesting aspects of these results is that, in fact, measuring the relaxation at a given temperature is completely equivalent to measure it at a different temperature but shifting the observation time window according to the law $\svar$. In this sense, the method enables us to obtain the relaxation curve at a certain temperature, in a time range that is not experimentally accessible, by simply dividing the $\svar$ axis by this temperature.In this particular example, we can obtain the relaxation curve at the lowest measured temperature of 1.8 K at times as large as $10^{119}$ s, which is obviously an experimentally inaccessible time. For sample 1, where the highest temperature that we have measured was 37 K, we are observing the relaxation curve corresponding to 2 K at times as large as $10^{173}$ s. 
According to expression \ref{Viscodis}, the viscosity is a function of the scaling variable $\svar$ and is proportional to $T$, so if we plot $S/T$ as a function of $\svar$ at low enough temperatures, the resulting curve will be the energy distribution function of the sample. As has already been mentioned, it is not possible to obtain an experimental relaxation curve covering enough time decades to map the whole energy barrier distribution at any temperature. 
However, the master curve can be used to obtain the relaxation curve at the lowest measured temperature extrapolated to experimentally unaccessible times. We have obtained $S$ by making the numerical derivative of the master curve. The results are shown in Fig. \ref{Tlntsam1_fig}, where we also show the derivative of the thermoremanence relative to the saturation magnetization versus the temperature ($dM_r/dT$) \cite{Linderothjm93} which is known to be proportional to the distribution function of blocking temperatures \cite{Chantrellie91} and consequently to the distribution function of energy barriers. The results obtained by the two methods are in very good agreement.

One way to check the self consistency of the method is to compare $f(E)$ as obtained from the viscosity with the one obtained by fitting the master relaxation curve to Eq. \ref{Mt} assuming only a logarithmic-linear distribution of energy barriers. The resulting distribution function is shown
in Fig. \ref{Tlntsam1_fig} in dashed lines.

\section{Models of interacting 1D chains of NPs}  

\begin{figure}[htbp]
\centering
\includegraphics[width= 0.7\textwidth]{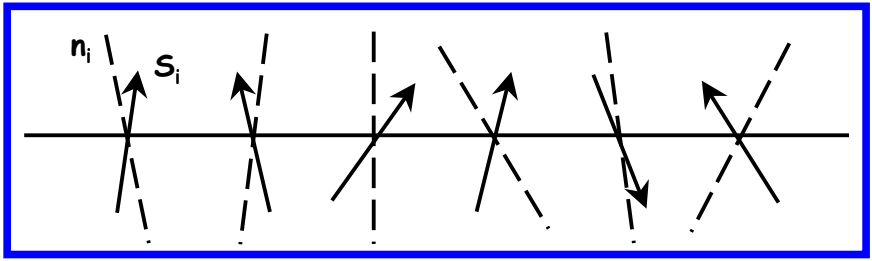}
\caption{1D chain of spins ${\bf S}_i$ with random anisotropy directions
${\bf n}_i$.}
\label{T7_1DPart_fig}
\end{figure}

In order to study the effect of long range interactions on the dynamics of spin systems, we will start with the simplest model capturing the essential physics of the problem.
Let us consider a linear chain of $N$ Heisenberg spins ${\bf S}_i$ $i=1,\dots, N$, each one representing a monodomain particle with magnetic moment ${\bf \mu}_i=\mu{\bf S_i}$. As indicated in Fig. \ref{T7_1DPart_fig}, each spin has uniaxial anisotropy pointing along the direction ${\bf \hat n}_i$, which may be the same for all and is oriented at random, and anisotropy constants, $K_i$, distributed according to a distribution function $f(K)$. An external magnetic field $\bf H$ may act on all the spins with the same value and pointing along the direction perpendicular to the chain. For simplicity, we will consider that particles have no internal structure, so that the only interaction taken into account will be the dipolar long-range interaction. The corresponding Hamiltonian is therefore:
\beq 
{\cal H}=-\sum_{i=1}^{N} \lbrace 
K_i ({\bf S}_i \cdot{\bf \hat n}_i)^2
+{\bf S}_i \cdot {\bf H} \rbrace 
+g\sum_{i=1}^{N} \sum_{j\neq i}^{N}
\left\{\frac{{\bf S}_i \cdot {\bf S}_j}{r_{ij}^3}-
3\frac{({\bf S}_i \cdot {\bf r}_{ij})({\bf S}_j \cdot {\bf r}_{ji})} 
{r_{ij}^5}\right\}
\ ,
\label{Hdip1}
\eeq
where $g=\mu^2/a^3$ characterizes the strength of the dipolar energy and $r_{ij}$ is the distance separating spins $i$ and $j$, $a$ is the lattice spacing, here chosen as $1$. 

\begin{figure}[htbp]
\centering
\includegraphics[width= 0.7\textwidth]{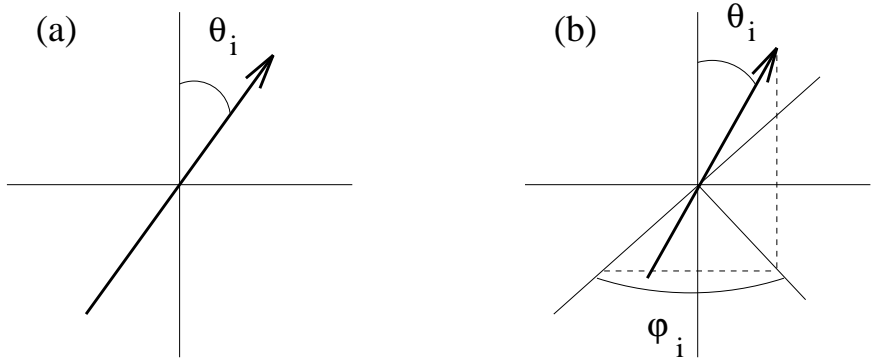}
\caption{The two 1D models considered. (a) Model I: 1D chain of planar spins,
(b) Model II: 1D chain of 3D spins.}
\label{T7_1DModel_fig}
\end{figure}

We will consider systems with either uniform anisotropy $K_i=K_0=1$ or with
a lognormal distribution of anisotropies, $f(K)$, of width $\sigma$ and centered 
at $K_0= 1$ . We will also study the effect of the random orientation of anisotropy 
axis $\bf n_i$ and compare this case with the case of aligned particles.  
Moreover, two different spin models will be condidered:
\begin{enumerate}
\item[a.] {\bf Model I} Spins in the same ($x-z$) plane with $\varphi_i =0$ 
characterized only by the angle $\theta_i$ (see Fig. \ref{T7_1DModel_fig}a). 
\item[b.] {\bf Model II} Spins with 3 dimensional orientations characterized 
by the spherical angles $\theta_i, \varphi_i$ (see Fig. \ref{T7_1DModel_fig}b).
\end{enumerate}
Besides the dimensionality of the spin vector, the difference between the two models lies on the fact that for Model I it is possible to write down an algorithm to find the exact values of the minima of the energy function and the energy barriers, whereas for Model II this becomes extremely difficult since a new degree of freedom comes into play.
\begin{figure}[b]
\centering
\includegraphics[width= 0.5\textwidth]{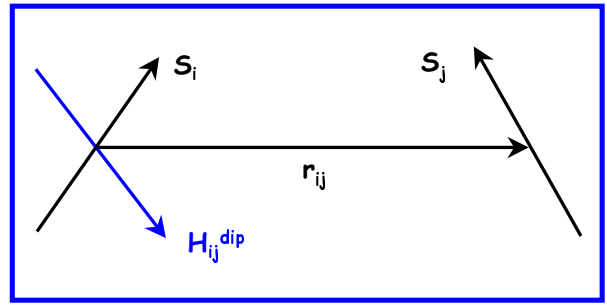}
\caption{Definition of the dipolar field ${\bf H}^{dip}_{ij}$ generated by the spin $S_j$ on the spin $S_i$.}
\label{T7_Particles_Hdip_fig}
\end{figure}

We will consider periodic boundary conditions along the chain, so that the restriction
\beq
{\bf S}_1= {\bf S}_N
\eeq
will be imposed in the simulations. In this way, we get rid of the possibility of spin reversal at the boundaries of the system because of the reduced coordination there. For this kind of boundary condition, the evaluation of dipolar fields for two or three dimensional spins would require replicating the system with several identical copies in order to minimize the rounding-off errors caused by the finite size of the system, but for our 1D model this will not be necessary. 

The effect of the dipolar interaction can be more easily understood if we define the dipolar fields acting on each spin $i$ (see Fig. \ref{T7_Particles_Hdip_fig}) 
\beq
{\bf H}^{dip}_i=-\sum_{j\neq i}^{N} 
\left\{ \frac{{\bf S}_j}{r_{ij}^3}-3
\frac{({\bf S}_j \cdot {\bf r}_{ji}){\bf r}_{ij}}
{r_{ij}^5}\right\}
\eeq 
and so, by rewriting the dipolar energy as 
\beq
{\cal H}_{\rm dip}=-g\sum_{i=1,}^{N}{\bf S}_i \cdot {\bf H}^{dip}_i \ ,
\eeq
the total energy of the system can be expressed into the simpler form
\beq
{\cal H}= -\sum_{i=1}^{N} \lbrace K_i ({\bf S}_i \cdot{\bf \hat n}_i)^2-
 {\bf S}_i \cdot {\bf H}_i^{eff}) \rbrace \ .
\eeq

Now, the system can be thought as an ensemble of non-interacting spins feeling an effective field which is the sum of an external and a locally changing dipolar field ${\bf H}_i^{eff}={\bf H}+ {\bf H}^{dip}_i$. Note now that the first term in (\ref{Hdip1}) is a demagnetizing term since it is minimized when the spins are antiparallel, while the second one tends to align the spins parallel and along the direction of the chain. For systems of aligned Ising spins only the first term is non-zero and, consequently, the dipolar field tends to induce AF order along the direction of the chain (the ground state configuration for this case). However, for Heisenberg or planar spins, the competition between the two terms give rise to frustrating interactions, which can induce other equilibrium configuration depending on the interplay between anisotropy and dipolar energies.
\section{Computational details}
\label{Comput_details}
\subsection{Calculation of dipolar energies}
The long-range character of the dipolar interaction, due to the double sum over $N$ in (\ref{Hdip1}), makes the energy computation in the standard MC algorithm extremely costly in CPU time in comparison with the other local energy terms. So that one is almost forced to find some way to reduce the time spent in the calculation of the dipolar energy. The first thing one can think of is to cut-off the interaction range to a sufficiently large (but still small, let us say 10) number of neighbours with the hope that the contribution of the furthest spins will be negligible. But it turns out that, even in 1D, sums of the kind $\sum_{n=1}^{\infty}1/n^\alpha$ are very poorly convergent so that truncation may result in considerable rounding-off errors and also to the formation of artificial surface charges or magnetic poles at the truncation distance which lead to non-physical solutions. 

Another strategy is to keep the exact calculation but to find an algorithm that avoids, in some way, the longest part of the calculation, i.e. the evaluation of the double sum in (\ref{Hdip1}) at every MC trial jump. Our algorithmic implementation for the calculation of dipolar energies is based in the following considerations:
\begin{enumerate}

\item Let us first note that it is not necessary to recalculate the dipolar energy each time a spin flip attempt is tried during the MC procedure. By rewriting the dipolar energy as
\beq
{\cal H}^{dip}= g\sum_{i,j=1}^{N}
S^{\alpha}_i W_{ij}^{\alpha\beta} S^{\beta}_j \ ,
\eeq
where $W$ is the dipolar interaction matrix
\beq
W_{ij}^{\alpha\beta}=\frac{\delta_{\alpha\beta}}{r_{ij}^3} -
3\frac{\delta_{\alpha\gamma}\delta_{\beta\eta}r_{ij}^{\gamma} r_{ij}^{\eta}}
{r_{ij}^5}
\ ,
\eeq
that represents the dipolar interaction between spins $i$ and $j$. Then these quantities can be calculated before starting the MC part of the program and stored in an array for later use, as they only depend on the position of the spins along the chain and not on the particular spin value.  

\item With these quantities at hand, it is more convenient to work with the dipolar fields as they are simply 
\beq
{H_{dip}}_{i}^{\alpha}=\sum_{j=1}^{N} W_{ij}^{\alpha\beta} S^{\beta}_j \ .
\label{Hdipolar}
\eeq
The dipolar fields may be calculated once for the initial spin configuration  before entering the MC part of the simulation and stored in an array.

\item Therefore, in the MC algorithm, the dipolar energy of a spin can be calculated simply by multiplying the stored dipolar field by the value of the spin as
\beq
{E_{dip}}_i=-{\bf S}_i \cdot {{\bf H}_{dip}}_i \ .
\eeq

\item Finally, let us note that, only if the trial jump is accepted, the dipolar fields have to be updated by a specific subroutine as
\beq
{{\bf H}_{dip}^{new}}_i =  {{\bf H}_{dip}^{old}}_i- {W}_{ij} 
({\bf S}^{old}_j-{\bf S}^{new}_j) \  (i=1,\dots, N)\ ,
\eeq
where ${\bf S}_j$ is the spin that has been changed. This requires only $N$ evaluations instead of the double $N$ sum involved in the calculation of the dipolar energy. This strategy is particularly efficient at low temperatures, when the acceptance rate can be very low.
\end{enumerate}
Note that this implementation of the calculation of dipolar energies is not limited to 1D systems. It can be applied to spin systems in to two or three dimensions and independently of their spatial arrangement in the nodes of a regular lattice or in random positions in space as long as they remain fixed in space.  

\subsection{The Monte Carlo algorithm}

\begin{figure}[htbp]
\centering
\includegraphics[width= 0.5\textwidth]{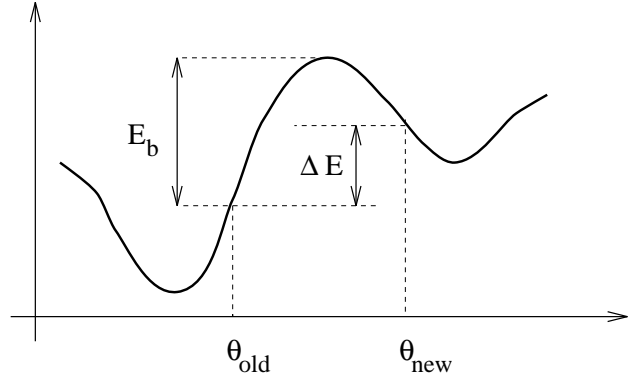}
\caption{Different choices of $\Delta E$ for the transition probability.}
\label{T7_Trialjump_fig}
\end{figure}

In Monte Carlo simulations of continuous spins, special care has to be taken in the way the attempt jumps are done, and in the way the energy difference $\Delta E$ appearing in the Boltzmann probability is calculated. They are mainly two choices for the dynamics of the MC procedure, independently of the way the attempt jumps are done: (a) either $\Delta E$ is directly calculated as the energy difference between the old ${\bf S}^{old}$ and the attempted ${\bf S}^{new}$ values of the spin or (b) $\Delta E$ is chosen as the energy barrier which separates ${\bf S}^{old}$ and ${\bf S}^{new}$ (see Fig. \ref{T7_Trialjump_fig}). Note that the second choice gives $\Delta E$'s that are higher than the first if there is an energy maximum separating the two states.
Each of the two models previously considered are suited to probe this two different choices: 
\begin{enumerate}
\item In Model I, the energy minima and the barriers separating them 
can be explicitly calculated, since, in this 
case, there is only one degree of freedom for each spin and its energy can be written as in 
the Stoner-Wohlfarth particle model of Section \ref{Model}
\beq
E_i=-K_i \cos^{2}(\theta_i-\psi_i)-H_{i}^{eff}\cos(\theta_i^h -\theta_i)  \ ,
\label{En1}
\eeq 
with an effective field which is the sum of the external and dipolar fields. Although the energy barriers cannot be analytically calculated for all the values of $\psi_i$ and $\theta_i^h$, it is not difficult to build up an algorithm that finds the minima and maxima of the energy function (\ref{En1}) and their respective energies.
Thus, we can build a MC algorithm that considers trial jumps only between orientations corresponding to energy minima randomly chosen with equal probabilities. The $\Delta E$ in the transition probability are always equal, in this case, to one of the actual energy barriers of the system.

\item In Model II, the 3D character of the spins and the dipolar field makes it difficult to devise an efficient algorithm to find the energy minima. Then, the trial jump must be done in this case to a random orientation inside a cone of aperture $\delta\theta$ around the current spin direction.
\end{enumerate}
Of course, when the MC simulation is used to simulate the evolution in time of a system, the link between the computer artificial MC step and real time will depend on how the $\Delta E$ is computed, but it is well-known that in either case, only in some specific situations \cite{Nowakprl00,Nowakann01} this correspondence can be stablished. 
\subsection{Dipolar fields in 1D} 

\begin{itemize}
	\item In a 1D chain of Ising spins with directions perpendicular to the line joining the spins, the ground state is AF, giving an energy $E_{\rm dip}^{AF}=-g/a^3$ for n.n spins. For spins pointing along the chain direction, the ground state is FM, giving now an energy $E_{dip}^{FM}=-2g/a^3$ for n.n. spins (see Fig. \ref{Halign_fig}).

	\item The local field felt by spins in a FM chain is 
		\begin{eqnarray}
					H_{\rm dip}^{FM}=-2g\ \zeta(3)= -2.4041\,g \,
		\end{eqnarray}
where $\zeta (s)$ is the Riemann's Zeta function \cite{Abramowitz}. While if the spins are AF ordered the dipolar fields are
		\begin{eqnarray}
				H_{\rm dip}^{AF}=\pm\ 2g\ \sum_{n=1}^{\infty}\frac{(-1)^n}{n^3}= \mp\ \frac{3}{2}\ 		           \zeta(3)= \mp\ 1.8031\,g \ .
		\end{eqnarray}
If the spins point along the chain, these values are doubled.
\end{itemize}
\section{Effective energy barrier distributions} 

Before starting to simulate the time dependence of the magnetization, we will study first the influence of dipolar interactions on the energy function of the particles. With this purpose, in this section we will compute the distribution of energy barriers of interacting systems with distributed properties in a similar way we did when we studied the effect of an homogeneous external magnetic field \cite{Iglesiasjap02}. As in that case, we can expect that the dipolar fields acting on each particle will modify the energy barrier distribution in absence of interactions. But now, since the local dipolar fields may vary depending on the spin configuration, the effective distributions will be different from that for homogeneous magnetic field.
We will concentrate on Model I, since in this case the energy barriers can be computed exactly and compared to those extracted from relaxation curves. Moreover, we will consider a system with a lognormal distribution ($\sigma=0.5$) of anisotropies $f(K)$ (see Eq. {\ref{logn}) and random anisotropy axis.

{\bf 1.}\ Let us first consider the case of a FM chain of spins pointing along the positive y axis ($\theta_i =0$). In Fig. \ref{T7_Eb_distrib_therma_fig} (left panel), we present the calculated energy barrier distributions for several values of the dipolar interaction $g$ and compare them with that for the non-interacting case $g=0$. The distributions have been obtained from histograms for a system with 40000 particles normalized to the total area of the distribution. 

In this case, the local dipolar fields on all the spins are exactly the same since periodic boundaries are assumed at the ends of the 1D chain. The dipolar field is given in this case by ${{\bf H}_{dip}^{FM}}_i=-2g\zeta(3)\hat y$. So the existence of one or two energy minima and the height of the energy barriers will be ruled by the ratio of the dipolar to anisotropy energies, since the reduced field reads now 
\begin{equation}
{h_{dip}^{FM}}_i= \zeta(3)\frac{g}{K_i} \ .
\end{equation}
Therefore, particles with ${h_{\rm dip}}_i< {h_{c}}_i(\psi_i)$ have two energy minima, while the rest will have only one. 

For small $g$ ($= 0.1$) there are slight changes on the $f(E_b )$ with respect to the non-interacting case. As it was the case for an external homogeneous field, the dipolar fields shift the peak of the distribution, while its shape is undisturbed. However, when increasing $g$, the smallest energy barriers of particles having the smallest $K$ start to disappear. This leads to the appearance of a peak at zero energy, an increase in the number of low energy barriers due to the reduction by the field, and to the appearance of a longer tail at high energies. As the dipolar interaction is increased further ($g=0.3, 0.4$) the original peak around $E_b \simeq 1$ is progressively suppressed as more barriers are destroyed, and a secondary distribution peaked at high energies appears as a consequence of barriers against rotation out of the field direction.
\begin{figure}[htbp]
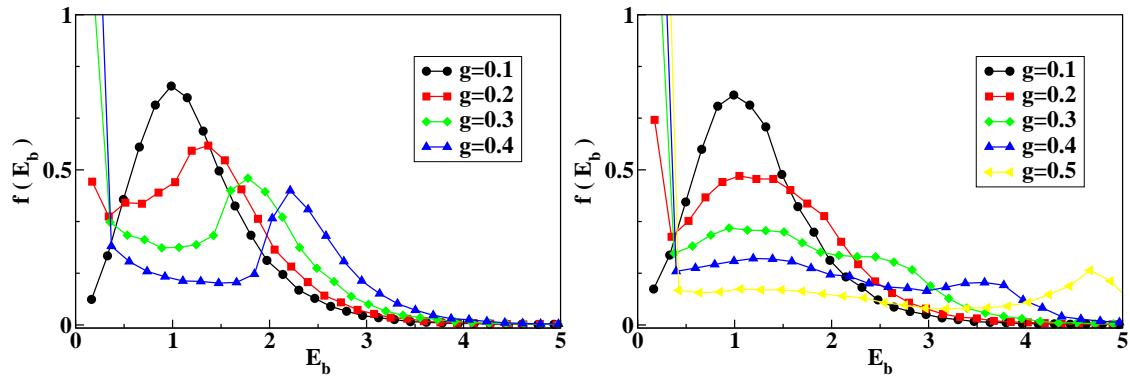

\centering
\includegraphics[width=0.45\textwidth]{Figures/T7_Eb_distrib_ali.eps}
\includegraphics[width=0.45\textwidth]{Figures/T7_Eb_distrib_therma.eps}
\caption{Energy barrier distributions $f(E_b)$ for a system with lognormal 
distribution of anisotropy constants ($\sigma=0.5$) and random anisotropy axes directions.
Left panel: spins pointing towards the y positive direction.
Right panel: spin configuration achieved after equilibration at $T=0$ in which spins have been replaced itereatively towards the nearest energy minimum direction starting from an initial FM configuration.
}
\label{T7_Eb_distrib_therma_fig}
\end{figure}

{\bf 2.}\ The previously analyzed configuration is highly metastable even at $T=0$ since, in general, the spin orientations are not along the energy minima. If the system is initially prepared this way by the application of a strong external field, for example, the particles will instantaneously reorientate their magnetizations so that they lie along the nearest minimum. This accommodation process occurs in a time scale of the order of $\tau_0$, much shorter than the thermal over-barrier relaxation times $\tau$. Therefore, in real experiments probing magnetization at time scales of the order of $1-10$ s ({\it i.e.} SQUID magnetometry), this will not be observed. In order to get rid of this ultra-fast relaxation during the first instants of the simulations, we submit the system to a previous equilibration process at $T=0$, during which the spins are consecutively placed in the nearest energy minima. Since the dipolar field after each of this movements changes on all the spins, the energy minima positions change continuously, but, after a certain number of MCS, the total magnetization changes become negligible and the system reaches a final equilibrated state. The effective energy barrier distributions after equilibration are displayed in the right panel of Fig. \ref{T7_Eb_distrib_therma_fig}. As we see, the high energy peaks observed in the FM configuration disappear almost completely after this process, indicating that they were due to the "misplacement" of the spins far from the local nearest minima. 
\section{Relaxation curves: $\svar$ scaling with interactions}
Here we would like to answer the following questions: how is the relaxation rate affected by dipolar interactions between the particles?, and is the $\svar$ scaling still accomplished even though the energy barrier landscape changes as time elapses in this case?. If so, then what is the meaning of the effective energy barrier distribution derived from the scaled relaxation curves?.
\subsection{Simulations of the time dependence of the magnetization}

Let us start with the simple case of a system with anisotropy distribution $f(K)$ and anisotropy axes distributed at random. As already mentioned in the previous section, if the system is initially prepared with all spins aligned in the field direction, the spins relax to the nearest minimum within a time of the order $\tau_0$ that is not usually accessible in most relaxation experiments. Therefore, the simulated relaxation will not be exactly as in typical experiments measuring the variation of the magnetization after saturation in a strong magnetic field. 
\begin{figure}[htbp]
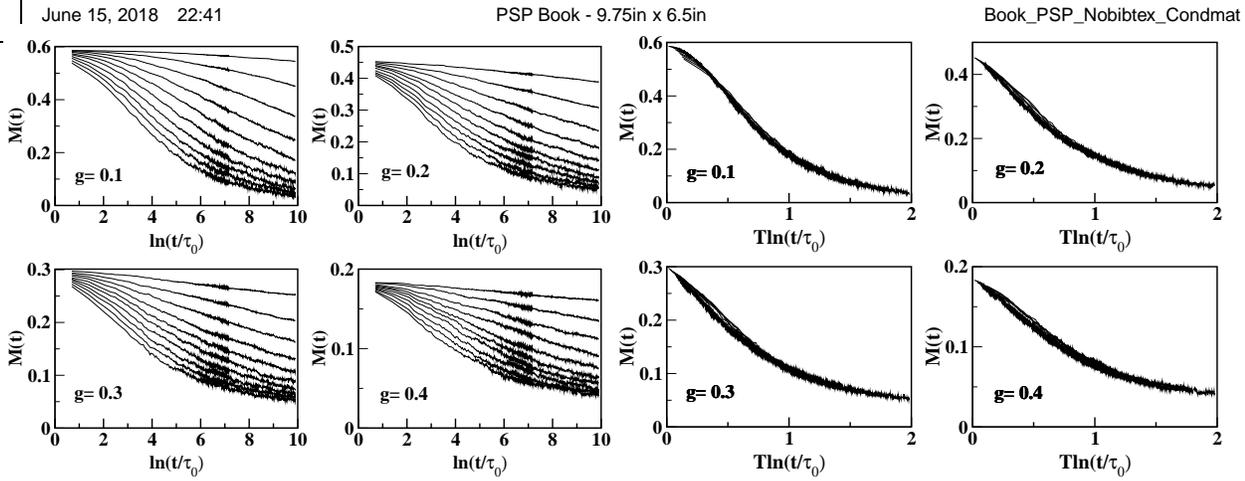

\centering
\includegraphics[width= 0.49\textwidth]{Figures/T7_Relax_log.eps}
\includegraphics[width= 0.49\textwidth]{Figures/T7_Relax_log_scaling.eps}
\caption{Left panels: Relaxation curves for several temperatures ranging from $T= 0.02$ (uppermost curves) to $T= 0.2$ (lowermost curves) in $0.02$ steps for a system of interacting particles with distribution of anisotropies $f(K)$ and random orientations. $g$ is the dipolar interaction strength. The initial state for all of them is the one achieved after the equilibration process described in the text. Right panels: Master relaxation curves corresponding to the relaxations shown in Fig. \ref{T7_Relax_log_fig} obtained by multiplicative scaling factor $T$.
} 
\label{T7_Relax_log_fig}
\end{figure}
We will run the simulations starting from the equilibrated states achieved by the previously described procedure. At non-zero $T$, thermal fluctuations will drive the system towards the equilibrium state, which for aligned particles would be AF. For a disordered system, however, the equilibrium state will have zero net magnetization. The final expected configuration will have neighbouring particles with almost parallel easy-axes in the $y$ direction with antiparallel spins and neighbouring particles with almost parallel easy-axes in the $x$ direction with parallel spins.  

The relaxation curves at different temperatures and values of the interaction $g$ are shown in Fig. \ref{T7_Relax_log_fig}. Temperature is measured in reduced units ($k_BT/K_0$), the chosen values of the interaction parameter range form the weak ($g=0.1$) to the strong ($g=0.5$) interaction regime.
We observe that the stronger the interaction, the smaller the magnetization of the initial configuration due to the increasing strength of the local dipolar fields that tend to depart the equilibrium directions from the direction of the anisotropy axis. Thus, we point out that, if one is to compare relaxation curves for different $g$ at the same $T$, they have to be properly normalized by the corresponding $m(0)$ value. As it is evidenced by the logarithmic time scale used in the figure, the relaxation is slowed down by to the intrinsic frustration of the interaction and the randomness of the particle orientations.
} 

More remarkable is the fact that the magnetization decay is faster the stronger the interaction is, which agrees well with the experimental results of Refs. \cite{Morupprl94,Batlleprb97,Garciaprb99} and also with other simulation works that model dipolar interactions by a mean-field \cite{Lyberatosjpd00,Lottisprl91,Dahlbergjap94,Matsonjap94}. However, at difference with these works, the quasi-logarithmic relaxation regime is only found in our simulations in the strong interaction regime, for short times, and within a narrow time window that depends on $T$. This can be understood because of the short duration of the relaxations in other works compared to ours, which were extended up to 10000 MCS, thus confirming the limitation of the logarithmic approximation to narrow time windows.

\subsection{$\svar$ scaling in presence of interaction.} 
Following the ideas exposed in Section \ref{Nonint}, we will try to analyze the relaxation curves at different temperatures according to the phenomenological approach of $\svar$ scaling. The underlying hypothesis of the method was that the dynamics of the system can be described in terms of thermal activation of the Arrhenius type over the effective local energy barriers induced by the interaction. Although one could think that this assumption is only valid in non-interacting particle systems, we would like to stress that the $\svar$ scaling approach was first successfully introduced in studies of spin-glasses \cite{Prejeanjpe80,Omarijpe84,Castaignjpe91}. Although it is true that the dipolar interaction, being long ranged, changes the energy barrier landscape in a dynamic way during the relaxation, this does not imply that the low $T$ relaxations will not scale. In fact, if this scaling is accomplished, it will give us information of the energy barriers that are effectively probed during the relaxation process, even if they keep on changing during the process.
\begin{figure}[htbp]
\centering
\includegraphics[width= 0.8\textwidth]{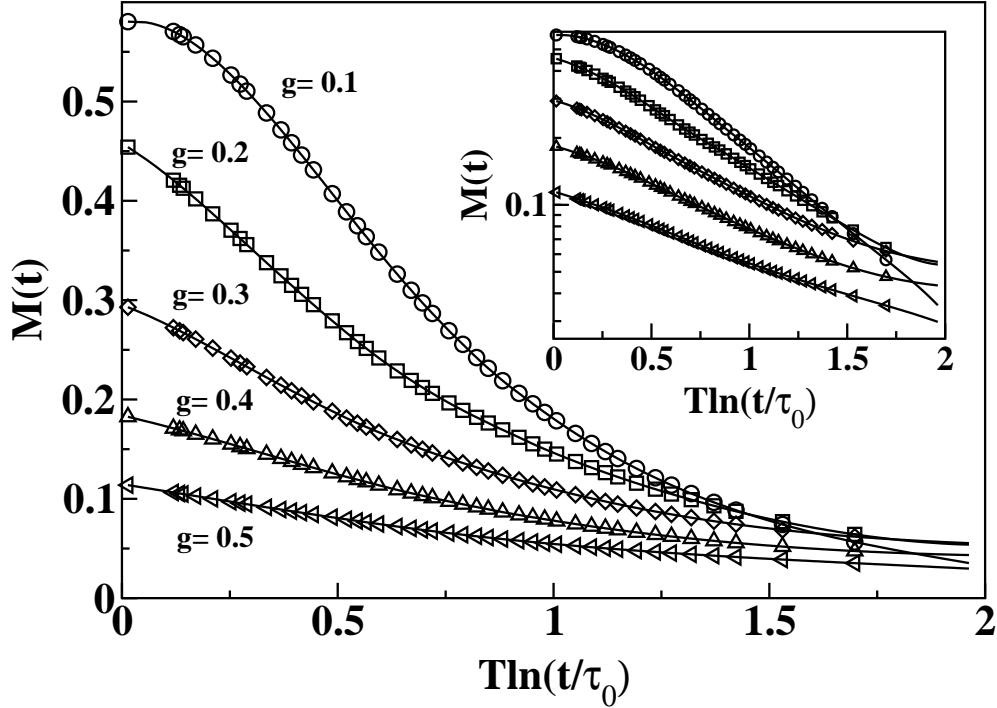}
\caption{Master relaxation curves for different values of the dipolar interaction strength $g$. Inset: the same curves in log-log plot in order to make evident the power-law behaviour of the relaxation at high values of $g$.
}
\label{T7_Masters_all_fig}
\end{figure}

The results of the master curves obtained from Fig. \ref{T7_Relax_log_fig} (left panels) by scaling the curves along the horizontal axis by multiplicative factors $T$ are presented in Fig. \ref{T7_Relax_log_fig} (right panels) for a range of temperatures covering one order of magnitude. 
First, we observe that, in all the cases, there is a wide range of times for which overlapping is observed. Below the inflection point of the master curve, the overlap is better for the low $T$ curves, whereas high $T$ curves overlap only at long times above the inflection point, as it was also the case in the non-interacting case. 
Moreover, it seems that scaling is accomplished over a wider range of $T$ the stronger the interaction is, whereas in the weak interaction regime, scaling is fulfilled over a narrower range of times and $T$. As we will explain later, this fact is due to the different variation of the effective energy barriers contributing to the relaxation in the two regimes.  

In order to see the influence of $g$ on the relaxation laws, in Fig. \ref{T7_Masters_all_fig}, we have plotted together the master relaxation curves for different values of the interaction parameter $g$ after a smoothing and filtering of the curves in Fig. \ref{T7_Relax_log_fig}. We can clearly see the qualitative change in the relaxation law with increasing $g$. In the weak interaction regime ($g=0.1,0.2$), the magnetization decays slowly to the equilibrium with an inflection point around which the decay law is quasi-logarithmic. In the strong interaction regime, however, the relaxation curves have always downward curvature with no inflection point. When plotted in a $\ln(M)$ {\it vs.} $\ln(t/\tau_0)$ scale they are linear (see Inset of Fig. \ref{T7_Masters_all_fig}), indicating a power-law decay of the magnetization with time, since the energy scale can be converted to time through the $\svar$ variable. 

This power-law behaviour has also been found by Ribas et al. \cite{Ribasjap96} in a 1D model of Ising spins and by Sampaio et al. \cite{Sampaioprb01,Sampaioprb96} in a Monte Carlo simulation of the time dependence of the magnetic relaxation of 2D array of Ising spins under a reversed magnetic field. It has also been observed experimentally in arrays of micromagnetic dots tracked by focused ion beam irradiation on Co layer with perpendicular anisotropy \cite{Hyndmanjm02,Aignprl98}.
\section{Evolution of $f_{\rm eff}(E_b)$ and of dipolar fields}
\begin{figure}[htbp]
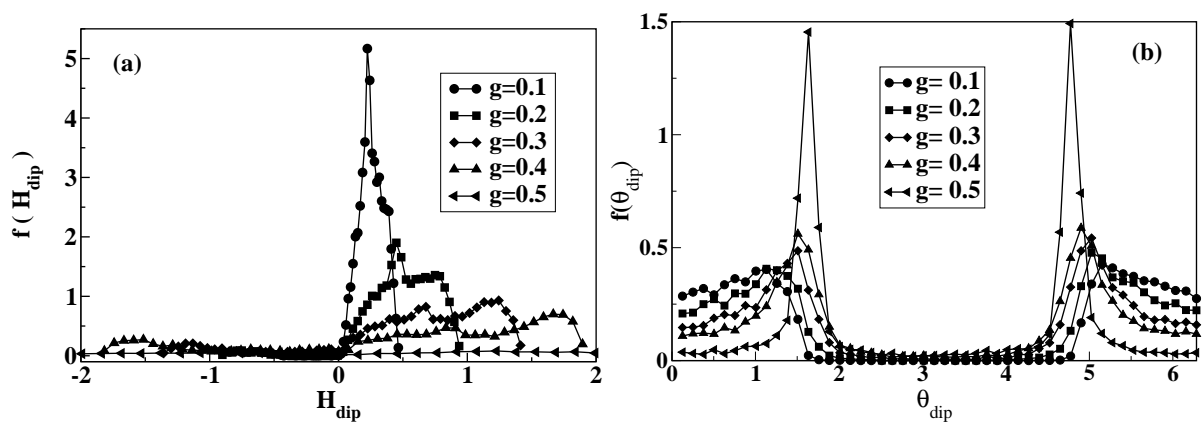

\centering
\includegraphics[width= 0.48\textwidth]{Figures/T7_Hloc_distrib_therma.eps}
\includegraphics[width= 0.48\textwidth]{Figures/T7_Tloc_distrib_therma.eps}
\caption{Left panel: Initial distribution of dipolar fields for a system of particles equilibrated at $T=0$ for different values of the interaction parameter $g$. All the magnetic moments are pointing along the local energy minima. Right panel: Initial distribution of dipolar field angles for the same system.
}
\label{T7_Hloc_distrib_therma_fig}
\end{figure}

In order to gain some insight on what are the microscopic mechanisms that rule the relaxation processes in the weak and strong interaction regimes, we will examine how the distribution of energy barriers and the dipolar fields change during the relaxation process. 
The initial distributions of energy barriers have already been shown in Fig. \ref{T7_Eb_distrib_therma_fig}, but since the starting configurations are not uniform, it is not easy to infer its microscopic origin. 
Let us notice that the distribution of dipolar fields [Eq. (\ref{Hdipolar})] does not depend on the anisotropy or easy-axis directions of the particles, so that it is only sensitive to the spin orientations and their positions in the lattice.
For this purpose, it turns useful to perform histograms of the strengh of the dipolar fields felt by all the spins at different values of $g$. 
We show the results in Fig. \ref{T7_Hloc_distrib_therma_fig}, where the dipolar fields having a component in the negative $y$ direction have been given a negative sign. 
This means that the local field is pointing in the opposite direction with respect to the original spin orientation, which was along the positive $y$ axis. Therefore, the existence of negative dipolar fields indicates a higher probability for the spin to jump towards the equilibrium state. 

For weak interaction ($g= 0.1$), the initial $f(H_{\rm dip})$ are strongly peaked at a value which is very close to the dipolar field for a FM configuration $H_{\rm dip}^{FM}=-2.0411381632g$, and there are very few negative dipolar fields. This indicates that the equilibrated configuration is not far from the initial FM one. In this case, since the dipolar fields are weak, the spins will point near the anisotropy axis direction since the energy minima and the energy barriers between them do not depart very much from the non-interacting case. This is also corroborated by the shape of $f(E_b)$, which resembles that for $g=0$.
 
However, in the strong interaction regime, the local fields start to destroy the energy barriers of the particles with lower $K$, and so negative dipolar fields are numerous because of the particles that have rotated into the field direction. There are still positive fields, but now the peak due to collinear spins blurs out with increasing $g$ at the same time that a second peak, centered at higher field values, starts to appear and finally swallows the first (see the case $g=0.5$). This last peak tends to a value equal to $H_{\rm dip}^{AF}= \mp\ 4.808\,g$, that would correspond to FM alignment along the chain. 

\begin{figure}[htbp]
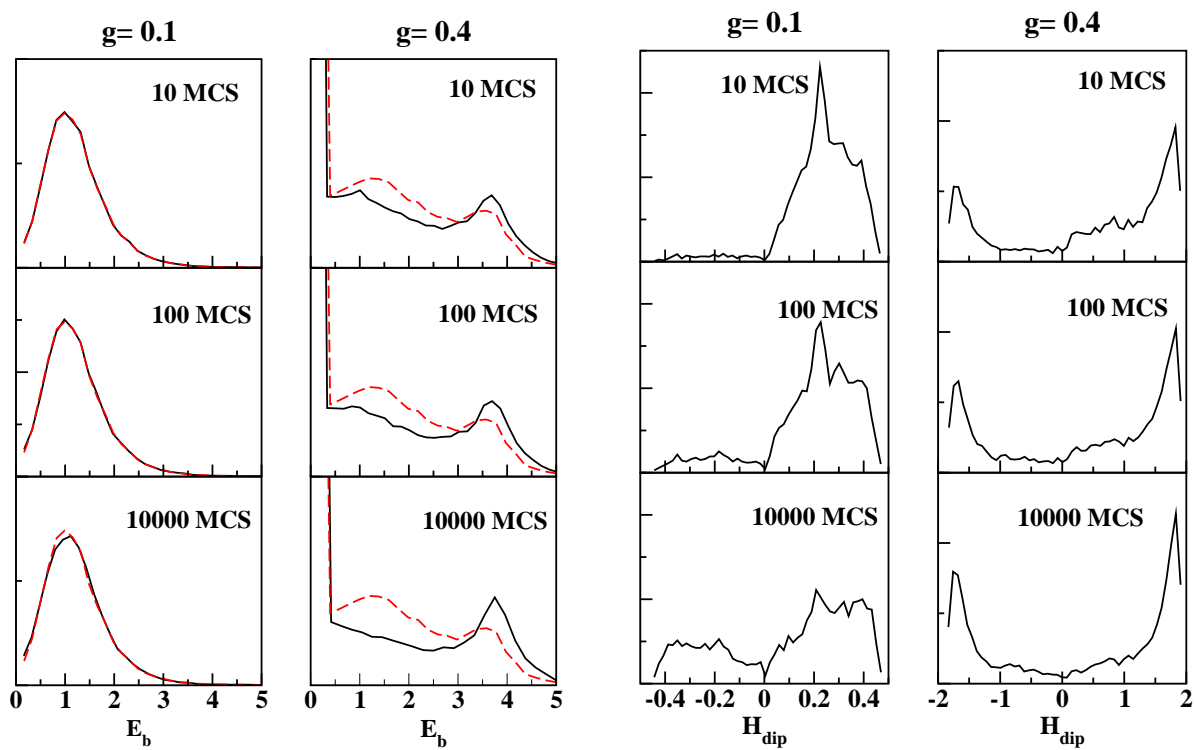

\centering
\includegraphics[width= 0.45\textwidth]{Figures/T7_Eb_distrib_finals.eps}
\hspace{0.05\textwidth}
\includegraphics[width= 0.45\textwidth]{Figures/T7_Hloc_distrib_finals.eps}
\caption{Left columns: Evolution with time of the energy barrier histograms during the relaxation process at $T= 0.1$. The initial distribution is shown in dashed lines.
Right columns: Time dependence of the distribution of dipolar fields during the relaxation process.
}
\label{T7_Eb_distrib_finals_fig}
\end{figure}

All these features are also supported by the distributions of dipolar field angles (right panel in Fig. \ref{T7_Hloc_distrib_therma_fig}). Now we can understand how the initial stages of the relaxation proceed. In order to gain a deeper insight into the microscopic evolution of the system during the relaxation, the histograms of energy barriers of intermediate configurations have been recorder at different MC steps. In Fig.\ref{T7_Eb_distrib_finals_fig}, the time evolution of the energy barriers separating the occupied state of each spin from the other allowed state is shown for a relaxation at an intermediate temperature $T=0.1$. In  \ref{T7_Eb_distrib_finals_fig}b, we have also kept track of the time dependence of the dipolar field histograms $f(H_{\rm dip})$. 

These evolutions are markedly different in the two interaction regimes.
In the weak interaction regime, the relaxation is dominated by anisotropy barriers, so that the distributions are similar to the non-interacting case. 
As time elapses, particles with the lowest energy barriers relax towards a state with higher energy barriers. However, although during the relaxation process the energy barriers change locally, this change is compensated by the average over the anisotropy distribution and random orientations of the easy-axes. 
Thus, the global  $f(E_b)$ does not change significantly as the system relaxes, although the final configuration is much more disordered than the initial one. 
In spite of this, the distribution of dipolar fields, which is more sensitive to the local changes in spin configuration, presents evident changes with time. As relaxation proceeds, the high peak of positive $H_{\rm dip}$ progressively flattens, since it corresponds to particles whose magnetization is not pointing along the equilibrium direction. Particles that have already relaxed, create dipolar fields in the negative direction which are reflected in a subdistribution of negative $H_{\rm dip}$ of increasing importance as time evolves. Near the equilibrium state of zero magnetization, the relative contribution of positive and negative fields tend to be equal, since, in average, there are equal number of "up" and "down" pointing spins. 

In the strong interaction regime, dipolar fields are stronger than anisotropy fields for the majority of the particles, even at the earlier stages of the relaxation process. 
As time elapses, the number of small energy barriers, corresponding to the particles with smaller anisotropies, continuously diminishes as they are overcame by thermal activation. 
When relaxing to their equilibrium state, now closer to the dipolar field direction, the particles with initially small $E_b$ give rise to higher energy barriers and also higher dipolar fields on their neighbours. This is reflected in the increasingly higher peak in the $f(E_b)$ that practically does not relax as time elapses, causing the final distribution to be completely different from the initial one. 
What is more, as more particles relax, more particles feel an $H_{\rm dip}>H_{\rm anis}$ and, therefore, higher $E_b$ for reversal against the local field. This leads to faster changes in the dipolar field distribution and also is at the origin of the power-law character of the relaxations. Equilibrium is reached when $f(H_{\rm dip})$ presents equal sharp peaked contributions from negative and positive fields, since in this case there will be an equal number of particles with magnetizations with positive and negative components along the $y$ axis.
\section{Effective energy barrier distributions from $\svar$ scaling}
\begin{figure}[htbp]
\centering
\includegraphics[width= 0.5\textwidth]{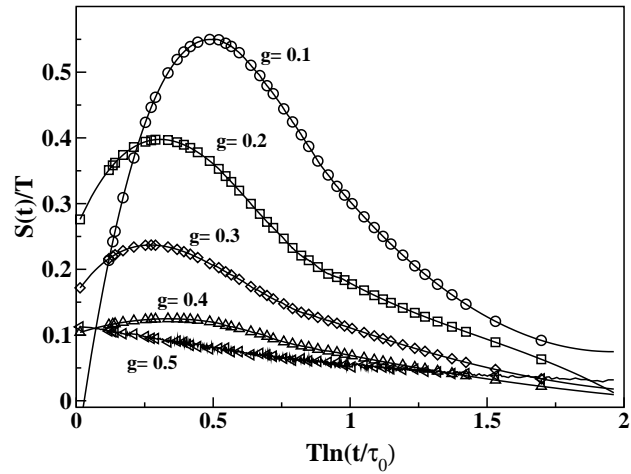}
\caption{Derivatives of the master relaxation curves of Fig. \ref{T7_Masters_all_fig} for different dipolar interaction strengths $g$.
}
\label{T7_Derivades_master_fig}
\end{figure}

In section \ref{Nonint}, we described a method to obtain an effective distribution of energy barriers from the master curves, showing that they can be obtained by performing the logarithmc time derivative of the master curves. The resulting effective energy barrier distributions obtained from the master curves in Fig. \ref{T7_Masters_all_fig} are given in Fig. \ref{T7_Derivades_master_fig}. It is worth remembering that these are not the real time evolving energy barrier distributions. Instead, they represent time independent distributions giving rise to the same relaxation curves obtained in the scaling regime. At difference with the non-interacting cases analyzed in previous chapters, these curves do not match the energy barrier distribution. The microscopic information given by them will be clarified in what follows.

For weak interaction ($g=0.1$), the effective distribution of energy barriers has essentially the same shape as for the non-interacting case. However, the distribution narrows as $g$ increases up to a value where almost zero barriers start to appear, and the mean effective barrier is shifted towards lower values of the scaling variable. In some sense, this resembles the situation for the non-interacting system in an external magnetic field, in which this shift was associated to the decrease of the energy barriers for rotation towards the field direction. 

When entering the strong interacting regime, an increasing number of low energy barriers appear that finally change the shape of the effective distribution into a quasi-exponential dependence. This change in the effective distribution is of course due to the power-law behaviour of the relaxation law in the strong $g$ regime and, therefore, a genuine effect of the dipolar interaction. This striking behaviour have important consequences on the experimental interpretation of relaxation curves. As we already mentioned in Section \ref{Nonint}, if magnetic relaxation is analyzed in terms of viscosity ({\sl i.e.}, the slope of the logarithmic time dependence), it turns out that, since $S\sim T\, f[\svar]$, an energy distribution diverging as $f(E)\sim 1/E$ would give a constant viscosity at low $T$ that could be erroneously interpreted as an indication of quantum $T$ independent relaxation phenomenon. 

\begin{figure}[htbp]
\centering
\includegraphics[width= 0.4\textwidth]{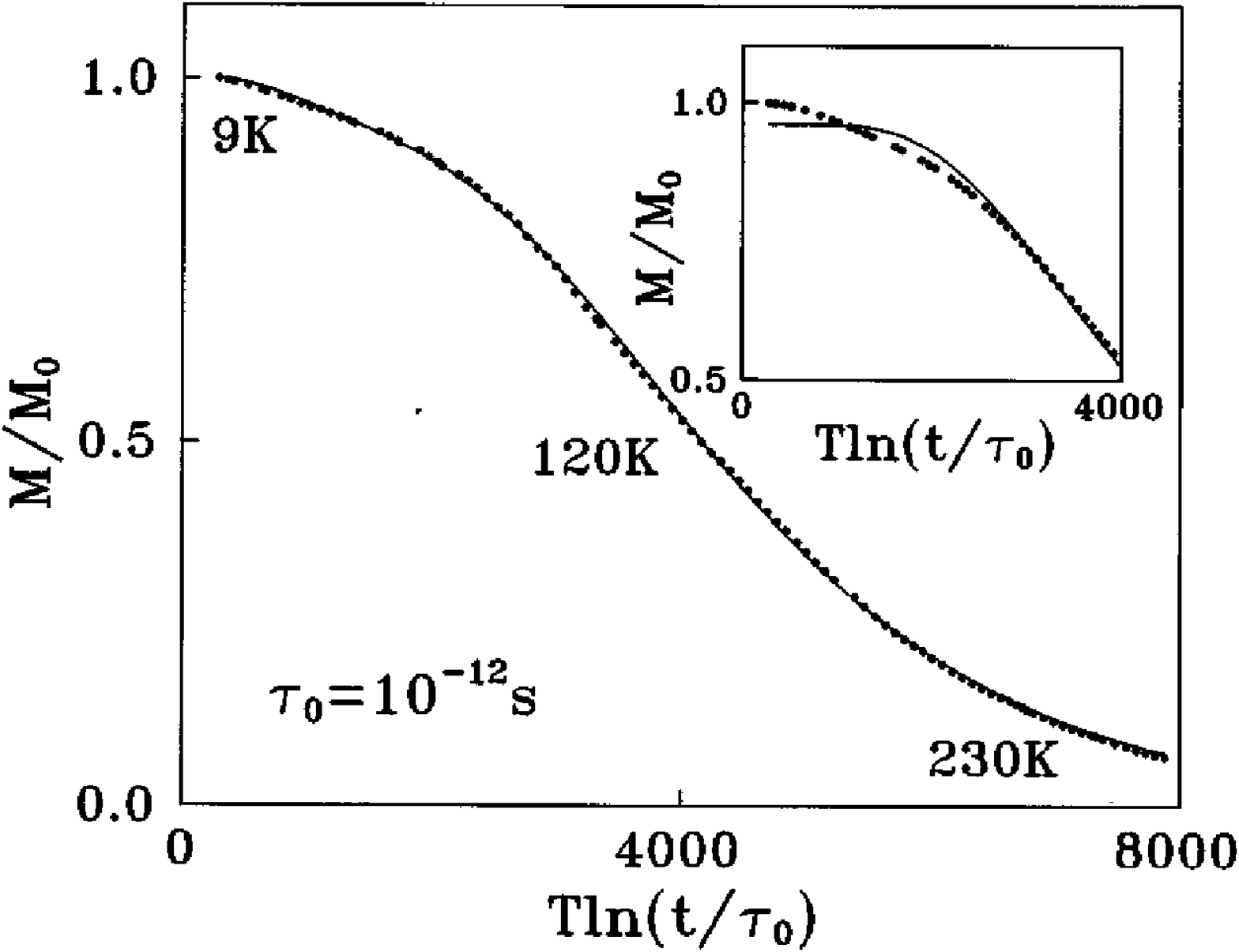}
\includegraphics[width= 0.4\textwidth]{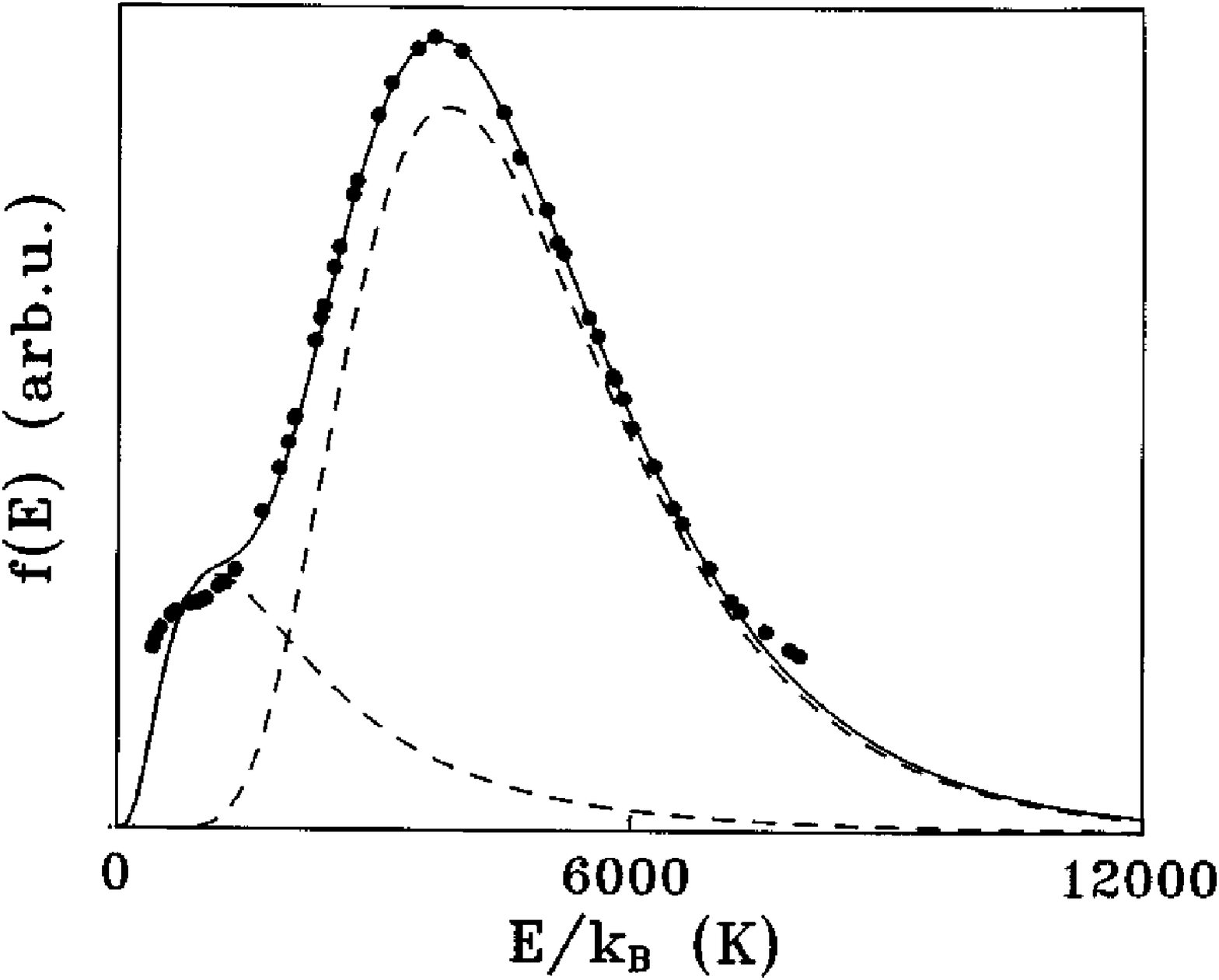}
\caption{Left panel: $M/M_0$ vs $\svar$ scaling with $\tau_0 =10^{-12}$ s for 27
temperatures within 9 and 230 K. Solid line represents the best fit of
data to Eq. (\ref{Dlogn}) considering two log-normal distributions of energy
barriers.
Right panel: Energy barrier distributions obtained from the derivative of the experimental master curve with respect to the scaling variable (filled circles).
The solid line indicates the fitted distribution, the dashed lines are the two log-normal subdistributions.
}
\label{T7_Montse_PRB97_mastercurve_fig}
\end{figure}

\begin{figure}[htbp]
\centering
\includegraphics[width= 0.5\textwidth]{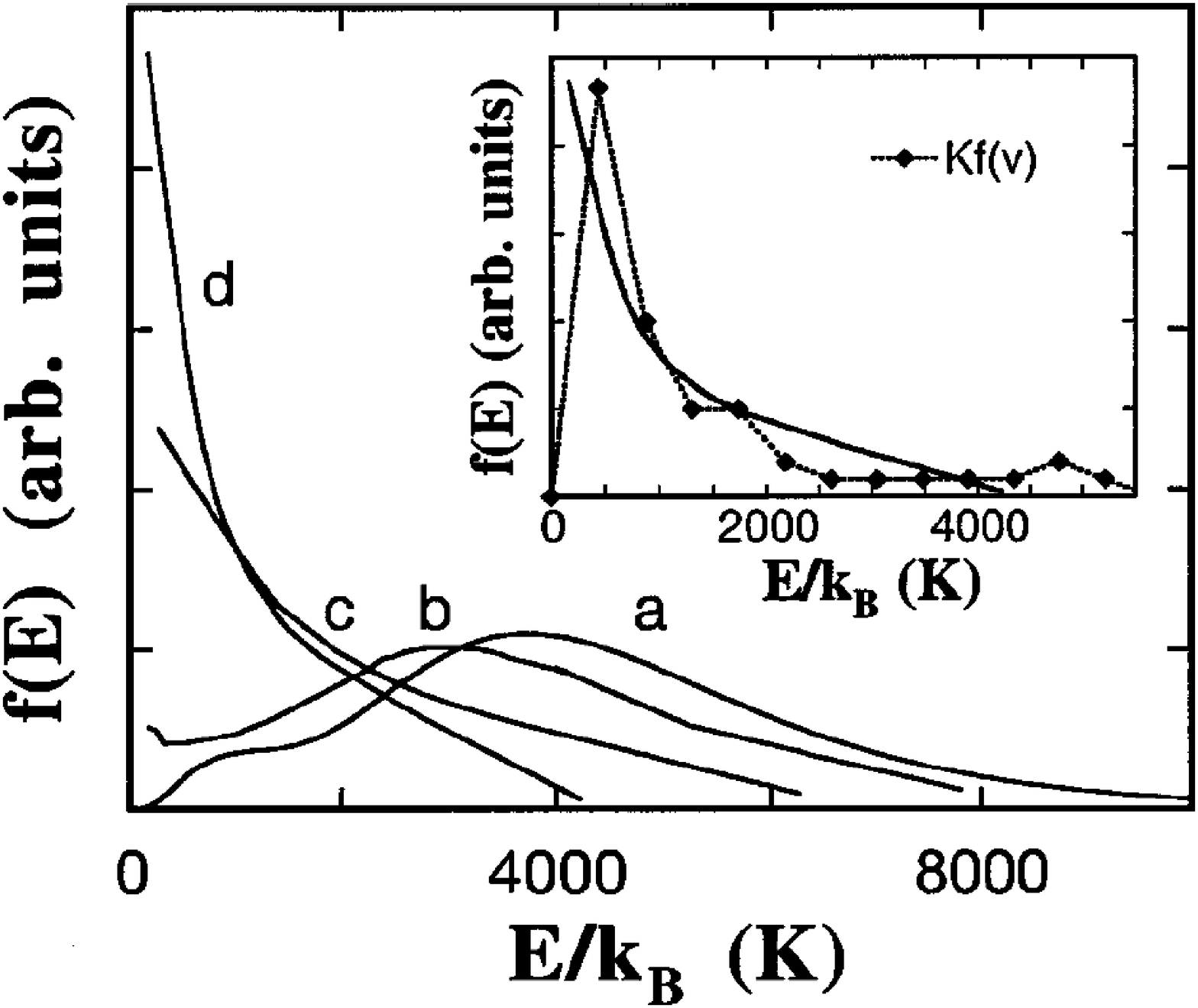}
\caption{Effective energy barrier distributions obtained from relaxation measurements in a Ba ferrite fine particle system 
after FC the sample at 200 Oe (a), 500 Oe (b), 10 kOe (c), and 50 kOe (d). $M_0$ is an arbitrary normalization factor.}
\label{T7_Montse_PRB99_derivadesmaster_fig}
\end{figure}

This change of behaviour in the effective energy barrier distributions and evidence of $\svar$ scaling of the relaxation has been observed experimentally in ensembles of Ba ferrite fine particles \cite{Batlleprb97}. The relevance of demagnetizing interactions in this scample was established by means of Henkel plots at different $T$. The relaxation curves of the thermoremanent magnetization for temperatures between $9$ and $230$ K can be scaled when plotted against the $\svar$ variable with $\tau_0= 10^{-12}$ s (see Fig. \ref{T7_Montse_PRB97_mastercurve_fig}, left panel). From the derivative of the master relaxation curve the distribution of effective energy barriers (see Fig. \ref{T7_Montse_PRB97_mastercurve_fig}, right panel) was obtained and fitted a sum of two log-normal distributions. The addition low energy barrier contribution to the energy barrier distribution can be associated to the demagnetizing interactions, since the other contribution centered at higher energies can be ascribed to the volume and anisotropy distributions. Moreover, when cooling the sample in different external magnetic fields before the relaxation process was recorded (see Fig. \ref{T7_Montse_PRB99_derivadesmaster_fig}), the authors found that, when increasing the cooling field, the effective distributions changed from a function with a maximum that extends to high enegies to a narrower distribution with a peak at much lower energy scales for high cooling fields. The effective distribution at high $H_{\rm FC}$, which was there argued to be given by the intrinsic anisotropy barriers of the particles, appears shifted towards lower energy values with respect to the anisotropy distribution as derived from TEM due to the demagnetizing dipolar fields generated by the almost aligned spin configuration induced by the $H_{\rm FC}$.  

\begin{figure}[htbp]
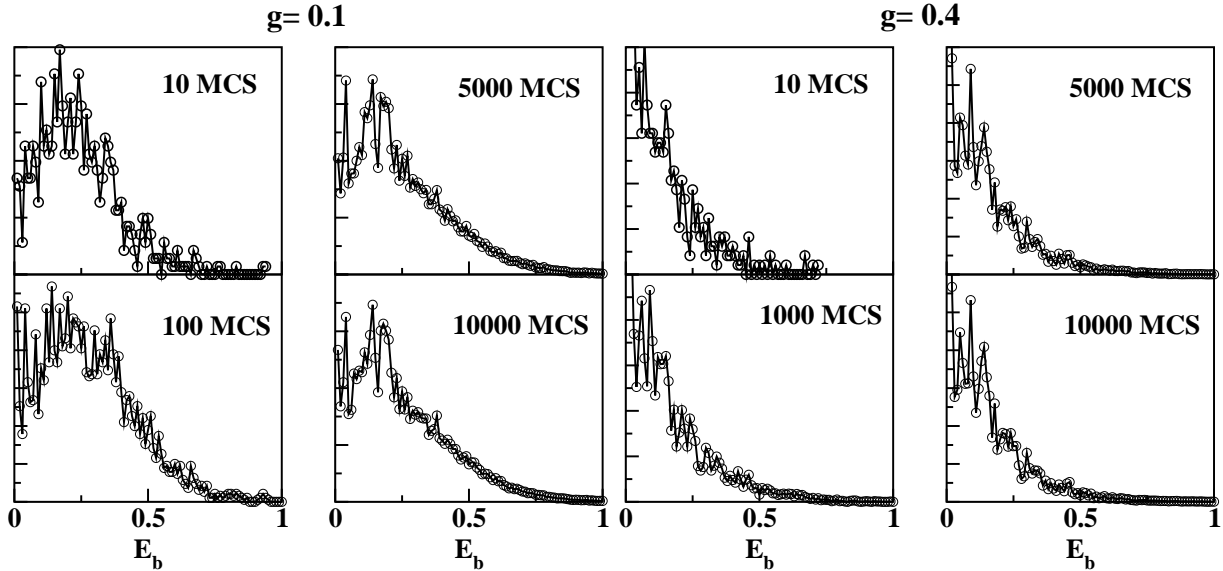

\centering
\includegraphics[width= 0.49\textwidth]{Figures/T7_Ebjump_g01_T010_a.eps}
\includegraphics[width= 0.49\textwidth]{Figures/T7_Ebjump_g04_T010_a.eps}
\caption{Cumulative histograms of the jumped energy barriers during the relaxation process. All the jumped energy barriers are taken into account. The temperature is $T= 0.1$. The value of the interaction parameter is $g= 0.1$ on left panels and $g= 0.4$ in right panels.
}
\label{T7_Ebjump_g01_T010_fig}
\end{figure}

As we have established that the effective energy barrier distributions derived form the master relaxation curves do not coincide with the real energy barrier distributions, now we will try to further clarify the meaning of these distributions. To this end, we have computed the cumulative histograms of energy barriers that have been really jumped during the relaxation processes. The corresponding results are presented in Fig. \ref{T7_Ebjump_g01_T010_fig} for systems in the weak and strong interaction regimes and $T= 0.1, 0.2$. As it is clear by comparison of the curves in this figure with those of Fig. \ref{T7_Derivades_master_fig}, although one could think that the derivative of the master curves collects jumped energy barriers of the order of $\svar$ as time elapses, the cumulative histograms overcount the number of small energy barriers at all the studied $T$ and $g$. This small energy barriers that are not seen by the relaxation correspond to the those jumped by the superparamagnetic (SP) particles. 

\begin{figure}[htbp]
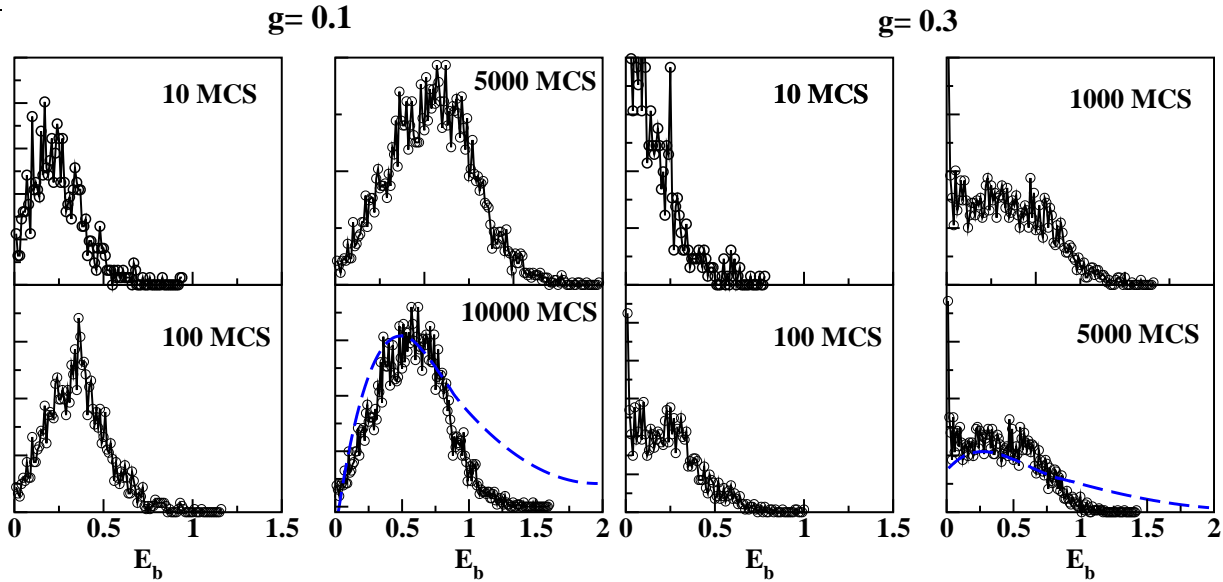

\centering
\includegraphics[width= 0.49\textwidth]{Figures/T7_Ebjumpnou_g01_T010_a.eps}
\includegraphics[width= 0.49\textwidth]{Figures/T7_Ebjumpnou_g03_T010_a.eps}
\caption{Cumulative histograms of the jumped energy barriers during the relaxation process. Only the $E_b$ jumped by particles that have not jumped up to time $t$ are taken into account. Symbols correpond to $T= 0.1$. The dashed lines stand for the derivatives of the master relaxation curves shown in Fig. \ref{T7_Derivades_master_fig}. 
The value of the interaction parameter is $g= 0.1$ on left panels and $g= 0.3$ in right panels.
}
\label{T7_Ebjumpnou_g01_T010_fig}
\end{figure}
In fact, when the cumulative histograms are computed by counting only the $E_b$ jumped by particles that have not jumped up to a given time $t$, the contribution of SP particles that have already relaxed to the equilibrium state is no longer taken into account. The histograms computed in this way are presented in Fig. \ref{T7_Ebjumpnou_g01_T010_fig}. There, we see that when only the energy barriers jumped by the blocked particles are taken into account, the resulting histograms at advanced stages of the relaxation process tend to the effective energy barriers derived from the master relaxation curves (dashed lines in the figure). The difference between both quantities at high energy values is due to the existence of very high energy barriers, that can only be surmounted at temperatures higher than the one considered here. 

\section{Hysteresis Loops}

Besides the time dependence of the magnetization, it is also interesting to study the effects of dipolar interactions on the hysteresis loops, since they give information about the reversal processes of the magnetization and are usual measurements in real samples. Since in this case we are not interested in how the system evolves with time but in finding the value of thermodynamic average of $M$ at a given $H$, we will implement the MC dynamics by choosing the Model II previously introduced in Sec. \ref{Comput_details}. In this way, the phase space is sampled more efficiently, minimizing computational and improving the quality of thermal averages. 
The studied system consists again of an ensamble of 10000 randomly oriented particles with log-normal distribution of anisotropy constants with $\sigma= 0.5$. We start the loop at high enough fields with a FM configuration and subsequently decrease the field in constant steps $\delta H=0.05$. At every field value, thermodynamic averages of the magnetization along the field direction are measured during a large but fixed number of MC steps.  
\begin{figure}[htbp]
\centering
\includegraphics[width= 0.7\textwidth]{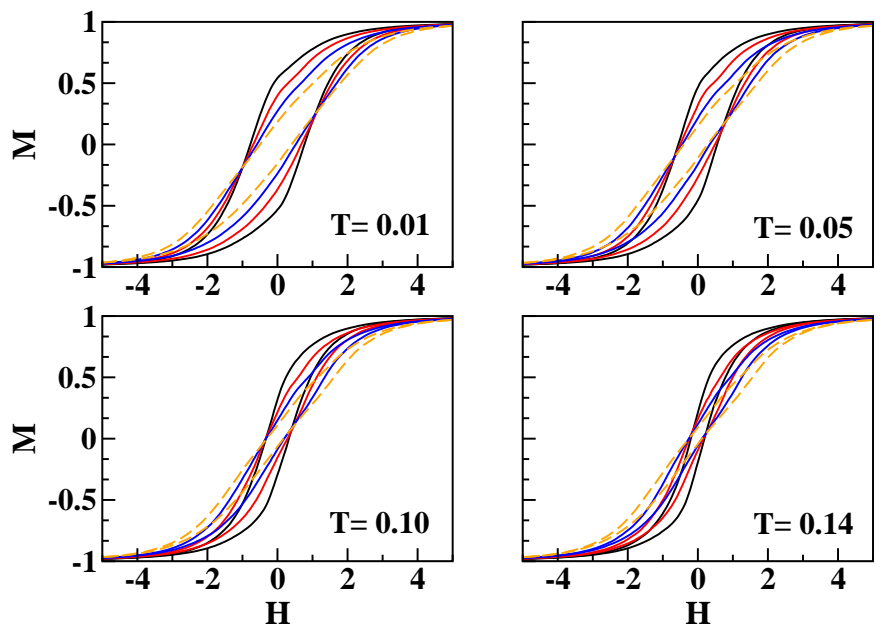}
\caption{Hysteresis loops for different values of the dipolar coupling $g= 0.1$ (black), $g= 0.2$ (red), $g= 0.3$ (blue), $g= 0.4$ (orange, discontinuous line). 
}
\label{T7_Hist_Loops_diffT_fig}
\end{figure}

In Fig. \ref{T7_Hist_Loops_diffT_fig}, we present the simulated hysteresis loops at different temperatures for values of $g$ ranging from $0.1$ to $0.4$. As it is apparent from the figures, the area of the hysteresis loops decreases with increasing temperature as expected. The loops become more elongated with increasing interaction resembling the ones for a system with frustrated interactions. The closure fields becomes higher and the system becomes harder as the it is more difficult to reach saturation.
\begin{figure}[htbp]
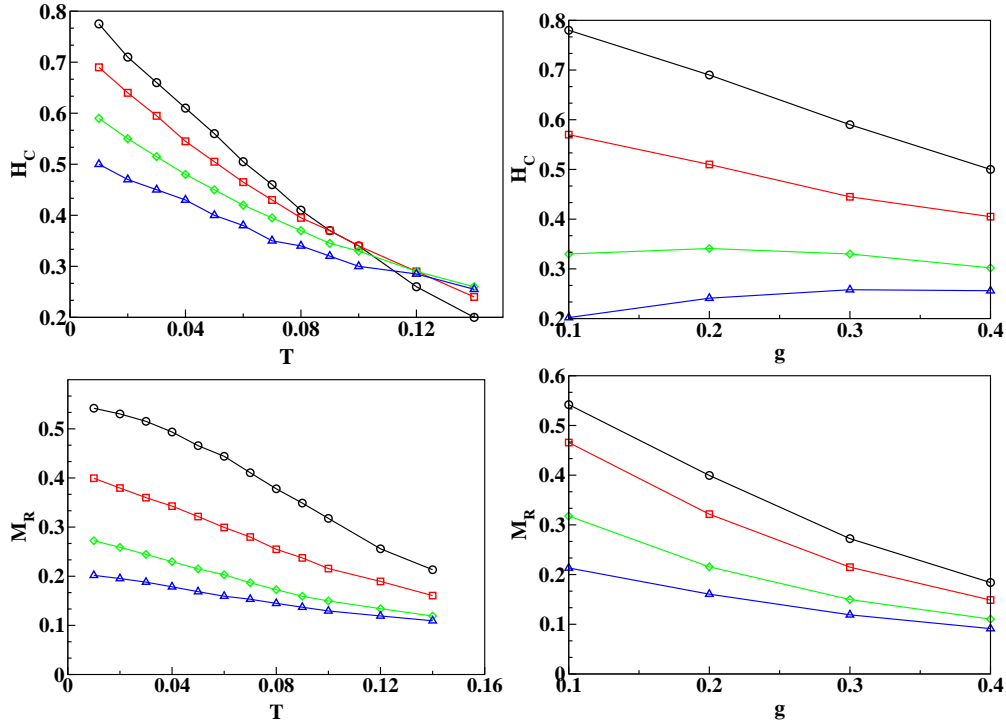

\centering
\includegraphics[width= 0.4\textwidth]{Figures/T7_H_C_T.eps}
\includegraphics[width= 0.4\textwidth]{Figures/T7_H_C_g.eps}
\includegraphics[width= 0.4\textwidth]{Figures/T7_M_Rem_T.eps}
\includegraphics[width= 0.4\textwidth]{Figures/T7_M_Rem_g.eps}
\caption{Left panels: Temperature dependence of the coercive field an remanent magnetization for $g= 0.1, 0.2, 0.3, 0.4$ (from the uppermost curve). Right panels: Dependence of the coercive field and remanent magnetization on the interaction parameter $g$ from temperatures $T= 0.01, 0.05, 0.1, 0.14$ (starting from the uppermost curve). 
}
\label{T7_H_C(T)_fig}
\end{figure}

The thermal dependence of the coercive field $H_c$ shown in Fig. \ref{T7_H_C(T)_fig} (upper pannels) for different values of the interaction parameter $g$ shows that, at low $T$, $H_c$ decreases linearly with increasing $T$ and also with increasing interactions. This observation is in agreement with experiments in interacting systems and simulations \cite{Kechrakosjm98,Xujap01}. However, at higher $T$ (see the $T= 0.1, 0.14$ curves in the right panel of Fig. \ref{T7_H_C(T)_fig}), $H_c(g)$ seems to have a maximum value at an intermediate $g$ value before starting to decrease for higher values of $g$. Further evidence of the influence of the frustration induced by the dipolar interaction is given by the thermal dependence of the remanent magnetization $M_R$. As shown in Fig. \ref{T7_H_C(T)_fig} (lower pannels), remanence values decrease with increasing interactions at all the considered $T$. Moreover, the thermal dependence of $M_R$ displays an inflection point at intermediate $T$ values, decaying smoothly towards zero for higher $T$, simiarly to experimental results for the thermoremanent magnetization of FeN ferrofluids of different concentrations \cite{Mamiyajap97}.

\section{Conclusions}
We have presented a review of a description of the long-time relaxation of the magnetization of nanoparticle ensembles based on a phenomenological approach to their dynamics that focuses on effective energy barriers. The approach is based on a scaling property of the relaxation curves at different temperatures that profits from the natural link between temperature and energy barriers through the Arrhenius law. We have also presented a way to obtain microcopy energy barrier distributions form the master relaxation curves obtained frmo the $\svar$ scaling. A proposal of extension of this methodology to interacting systems has been presented in detail as applied to a 1D chain of spins which has been shown to be valid in spite of the fact that energy barriers keep on changing through the relaxation process.

We have shown that the dipolar interaction induces a faster relaxation of the magnetization, changing the time dependence of the magnetic relaxation from quasi-logarithmic to a power law as g increases, due to the intrinsic disorder of the system and the frustration induced by the dipolar interactions. $\svar$ scaling of the relaxation curves at different $T$ is accomplished even in the presence of interaction.
From the obtained master curves, effective energy barrier distributions can be obtained, giving valuable information about the microscopic energy barriers and the change induced on them by the dipolar interaction which cannot be directly obtained experimentally. As the strength of the dipolar interaction $g$ increases, the effective energy barrier distribution shifts towards lower $E_b$ values and becomes wider, in qualitative agreement with experimental results. In spite of the dynamic change of the dipolar fields, the energy barrier distribution does not change appreciably during the relaxation due to the disorder induced by randomness of interaction. Morover, the results of simulations of the hysteresis loops display a decrease of the coercive field and remanance with increasing interaction in agreement with most of the experimental findings. 
The hysteresis loops of this system resemble those of frustrated systems, with elongated shapes and high closure fields. A reduction in the coercive field and the remanent magnetization with increaseing interaction is in agreement with experimental findings.

\section*{Acknowledgements}
The authors acknowledges funding of the Spanish MICINN through Grant project MAT2009-08667, European Union FEDER funds ("Una manera de hacer Europa"), Generalitat de Catalunya through Project 2009SGR876 through Project 2009SGR856, and CESCA  and CEPBA under coordination of C4 for supercomputer facilities.



\end{document}